\documentclass[twocolumn]{aastex631}

\accepted{February 17, 2022}

\submitjournal{ApJ}

\shorttitle{Measurement of $\mrm{^{26}Al}$ with COSI}
\shortauthors{Beechert et al.}
\graphicspath{{./}{figures/}}
\usepackage{xurl}
\usepackage{xspace}

 \usepackage{amsmath}

\newcommand{\mrm}[1]{\mathrm{#1}}
\newcommand{\nuc}[2]{$\mrm{^{#2}#1}$}
\newcommand{\flux}{$\mrm{ph\,cm^{-2}\,s^{-1}}$\xspace}
\newcommand{\vel}{$\mrm{km\,s^{-1}}$\xspace}

\begin{document}
%\linenumbers

\title{Measurement of Galactic \boldmath{\nuc{Al}{26}} with the Compton Spectrometer and Imager}

\correspondingauthor{Jacqueline Beechert}
\email{jbeechert@berkeley.edu}

\author[0000-0002-0128-7424]{Jacqueline Beechert}
\affiliation{Space Sciences Laboratory, UC Berkeley, 7 Gauss Way, Berkeley, CA 94720, USA}

\author[0000-0002-0552-3535]{Thomas Siegert}
\affiliation{Max-Planck-Institute for extraterrestrial Physics, Giessenbachstr. 1, 85748, Garching bei M{\"u}nchen, Germany}
\affiliation{Institut f{\"u}r Theoretische Physik und Astrophysik, Universit{\"a}t W{\"u}rzburg, Campus Hubland Nord, Emil-Fischer-Str. 31, 97074 W{\"u}rzburg, Germany}
\affiliation{Center for Astrophysics and Space Sciences, University of California, San Diego, 9500 Gilman Dr., La Jolla, CA 92093, USA}

\author[0000-0001-5506-9855]{John A. Tomsick}
\affiliation{Space Sciences Laboratory, UC Berkeley, 7 Gauss Way, Berkeley, CA 94720, USA}

\author[0000-0001-9067-3150]{Andreas Zoglauer}
\affiliation{Space Sciences Laboratory, UC Berkeley, 7 Gauss Way, Berkeley, CA 94720, USA}

\author[0000-0001-9567-4224]{Steven E. Boggs}
\affiliation{Center for Astrophysics and Space Sciences, University of California, San Diego, 9500 Gilman Dr., La Jolla, CA 92093, USA}

% \author[0000-0003-4087-1786]{Theresa J. Brandt}
\author[0000-0003-4087-1786]{Terri J. Brandt}
\affiliation{NASA Goddard Space Flight Center, Greenbelt, MD 20771, USA}

\author{Hannah Gulick}
\affiliation{Space Sciences Laboratory, UC Berkeley, 7 Gauss Way, Berkeley, CA 94720, USA}

\author[0000-0002-1757-9560]{Pierre Jean}
\affiliation{IRAP, 9 Av colonel Roche, BP44346, F-31028 Toulouse Cedex 4, France}

\author{Carolyn Kierans}
\affiliation{NASA Goddard Space Flight Center, Greenbelt, MD 20771, USA}

\author{Hadar Lazar}
\affiliation{Space Sciences Laboratory, UC Berkeley, 7 Gauss Way, Berkeley, CA 94720, USA}

\author{Alexander Lowell}
\affiliation{Space Sciences Laboratory, UC Berkeley, 7 Gauss Way, Berkeley, CA 94720, USA}

\author{Jarred M. Roberts}
\affiliation{Center for Astrophysics and Space Sciences, University of California, San Diego, 9500 Gilman Dr., La Jolla, CA 92093, USA}

\author[0000-0003-4732-6174 ]{Clio Sleator}
\affiliation{U.S. Naval Research Laboratory, Washington DC 20375, USA}

\author{Peter von Ballmoos}
\affiliation{IRAP, 9 Av colonel Roche, BP44346, F-31028 Toulouse Cedex 4, France}

% \collaboration{6}{(AAS Journals Data Editors)}

%% Note that the \and command from previous versions of AASTeX is now
%% depreciated in this version as it is no longer necessary. AASTeX 
%% automatically takes care of all commas and "and"s between authors names.

%% AASTeX 6.31 has the new \collaboration and \nocollaboration commands to
%% provide the collaboration status of a group of authors. These commands 
%% can be used either before or after the list of corresponding authors. The
%% argument for \collaboration is the collaboration identifier. Authors are
%% encouraged to surround collaboration identifiers with ()s. The 
%% \nocollaboration command takes no argument and exists to indicate that
%% the nearby authors are not part of surrounding collaborations.

%% Mark off the abstract in the ``abstract'' environment. 
\begin{abstract}
%
% 250 word limit
%
The Compton Spectrometer and Imager (COSI) is a balloon-borne compact Compton telescope designed to survey the 0.2--5\,MeV sky.
%
% COSI's energy resolution of $\sim$0.2\% at 1.8\,MeV, single-photon reconstruction, and wide field of view make it capable of imaging diffuse emission.
% %
% This facilitates studies of astrophysical nuclear lines, particularly the 1809\,keV $\gamma$-ray line from decaying Galactic \nuc{Al}{26}.
COSI's energy resolution of $\sim$0.2\% at 1.8\,MeV, single-photon reconstruction, and wide field of view make it capable of studying astrophysical nuclear lines, particularly the 1809\,keV $\gamma$-ray line from decaying Galactic \nuc{Al}{26}.
Most \nuc{Al}{26} originates in massive stars and core-collapse supernova nucleosynthesis, but the path from stellar evolution models to Galaxy-wide emission remains unconstrained.
In 2016, COSI had a successful 46-day flight on a NASA superpressure balloon.
Here, we detail the first search for the 1809\,keV \nuc{Al}{26} line in the COSI 2016 balloon flight using a maximum likelihood analysis.
We find a Galactic \nuc{Al}{26} flux of $(8.6 \pm 2.5) \times 10^{-4}$\,\flux within the Inner Galaxy ($|\ell| \leq 30^{\circ}$, $|b| \leq 10^{\circ}$) with 3.7$\sigma$ significance above background.
Within uncertainties, this flux is consistent with expectations from previous measurements by SPI and COMPTEL.
This analysis demonstrates COSI's powerful capabilities for studies of $\gamma$-ray lines and underscores the scientific potential of future compact Compton telescopes.
%
% In particular, COSI as a NASA Small Explorer satellite, slated for launch in 2025, will bring long-awaited clarity to the MeV sky. 
%
In particular, the next iteration of COSI as a NASA Small Explorer satellite has recently been approved for launch in 2025.
\end{abstract}

%% Keywords should appear after the \end{abstract} command. 
%% The AAS Journals now uses Unified Astronomy Thesaurus concepts:
%% https://astrothesaurus.org
%% You will be asked to selected these concepts during the submission process
%% but this old "keyword" functionality is maintained in case authors want
%% to include these concepts in their preprints.

%% From the front matter, we move on to the body of the paper.
%% Sections are demarcated by \section and \subsection, respectively.
%% Observe the use of the LaTeX \label
%% command after the \subsection to give a symbolic KEY to the
%% subsection for cross-referencing in a \ref command.
%% You can use LaTeX's \ref and \label commands to keep track of
%% cross-references to sections, equations, tables, and figures.
%% That way, if you change the order of any elements, LaTeX will
%% automatically renumber them.
%%
%% We recommend that authors also use the natbib \citep
%% and \citet commands to identify citations.  The citations are
%% tied to the reference list via symbolic KEYs. The KEY corresponds
%% to the KEY in the \bibitem in the reference list below. 

\section{Introduction} \label{sec:intro}
Aluminum-26 (\nuc{Al}{26}) is a radioactive isotope which traces the synthesis, dynamics, and incorporation of elements in the interstellar medium (ISM) of the Milky Way.
It decays to an excited state of Magnesium-26 (\nuc{Mg}{26}) with a half-life time of 0.715\,Myr.
The de-excitation of \nuc{Mg^*}{26} to its ground state emits a 1809\,keV $\gamma$-ray.
\nuc{Al}{26} lives long enough to decay into the ISM after it is ejected from its production sites.
This allows studies of the stellar conditions responsible for nucleosynthesis and the hot phase of the ISM. 

The High Energy Astronomy Observatory (HEAO-3) satellite reported the first detection of Galactic \nuc{Al}{26} in 1984 \citep{mahoney1984heao}.
In the 1990s, the Compton Telescope (COMPTEL) on board the Compton Gamma-Ray Observatory obtained the first images of \nuc{Al}{26} emission in the Milky Way.
COMPTEL revealed the diffuse emission in the Inner Galaxy ($|\ell| \leq 30^{\circ}$, $|b| \leq 10^{\circ}$) with a flux of $3.3 \times 10^{-4}$\,\flux.
Emission was also observed along the Galactic Plane, including the star-forming regions Cygnus, Carina, and Vela \citep{pluschke2001comptel}.
The 1.8\,MeV emission was found to be reminiscent of the population of massive stars, particularly those which are able to sustain ionized regions in the ISM \citep{knodlseder1999multiwavelength,knodlseder1999implications}.

The spectrometer SPI on board the International Gamma-ray Astrophysics Laboratory (INTEGRAL) satellite, launched by the European Space Agency in 2002, first detected the \nuc{Al}{26} line in 2006 with an Inner Galaxy flux of $(3.3 \pm 0.4) \times 10^{-4}$\,\flux \citep{diehl2006radioactive}.
Recent analyses of over a decade of data detect the line with 58$\sigma$ significance at $1809.83 \pm 0.04$\,keV and a full-sky flux of $(1.84 \pm 0.03) \times 10^{-3}$\,\flux \citep{moritz_thesis}.
The flux from the Inner Galaxy was found to be $(2.89 \pm 0.07) \times 10^{-4}$\,\flux \citep{siegert2017positron}.
SPI has also produced a 1.8\,MeV image largely consistent with that of COMPTEL \citep{Bouchet_2015}.
A recent review of the current understanding of \nuc{Al}{26} is provided by \citet{diehl2020steady}. %

Questions surrounding the influence of \nuc{Al}{26} on the formation of young solar systems also motivate characterization of its emission.
Observations of the Ophiuchus complex, for example, reveal flows of \nuc{Al}{26} originating from young stellar environments. 
Studying the dynamics of \nuc{Al}{26} in Ophiuchus may shed light on the formation of our own solar system and on the typical dynamics of its emission from stellar environments \citep{forbes2021solar}.
\citet{forbes2021solar} also suggest dominant emission of \nuc{Al}{26} in Ophiuchus by numerous supernovae rather than a single, large supernova event or Wolf-Rayet winds, although the contributions by several supernovae compared to Wolf-Rayet stars remain subject to considerable uncertainties.

These uncertainties, difficulties in simulating the dynamics of \nuc{Al}{26} emission, and evident disagreement between the structure of the ISM in simulated \nuc{Al}{26} maps and those from observations \citep{pleintinger2019comparing} require additional measurements.
Detailed observations of the 1.8\,MeV line and its spatial morphology are necessary to resolve the primary sources of \nuc{Al}{26} and its distribution throughout the ISM. 
In this work, we aim to establish the scientific potential of modern compact Compton telescopes in nucleosynthesis studies and thereby present a key proof-of-concept study for the Compton telescope satellite mission, COSI-SMEX, recently selected for launch as a NASA Small Explorer (SMEX) spacecraft in 2025\footnote{NASA press release: \url{https://www.nasa.gov/press-release/nasa-selects-gamma-ray-telescope-to-chart-milky-way-evolution}} \citep{tomsick2021compton,tomsick2019compton}.

Here, we use the balloon-borne precursor to COSI-SMEX, the Compton Spectrometer and Imager (COSI), a compact Compton telescope with excellent spectral resolution of 0.24\% FWHM at 1.8\,MeV.
Twelve high-purity cross-strip germanium semiconductor detectors (each $8 \times 8 \times 1.5\,\mrm{cm^3}$) are arranged in a $2 \times 2 \times 3$ array that measures photons between 0.2 and 5\,MeV.
The photon path through the detectors is reconstructed using the energy and three-dimensional position of each interaction \citep{boggs2000event}.
The incident photon is localized to a circle on the sky defined by the cosine of the first Compton scatter angle $\phi$ in the instrument.
A comprehensive review of calibrations and analysis principles of Compton telescopes is provided in \citet{zoglauer2021cosi}.
Six anti-coincidence cesium iodide (CsI) shields surrounding the four sides and bottom of the detector array constrain the wide $\sim$1$\pi\,\mrm{sr}$ field of view.
The shields suppress the Earth albedo radiation by actively vetoing $\gamma$-rays incident from below the instrument. 
The shield veto system reduces atmospheric background levels by $\sim$1--2 orders of magnitude above 1750\,keV.
Note that by installing these shields for atmospheric background rejection, we introduce the potential for instrumental activation of the shield materials. 
This activation can create background $\gamma$-ray lines in the data set which are accounted for empirically in the presented analysis.

In this work we demonstrate COSI's ability to perform high-resolution spectroscopy of astrophysical nuclear lines through the search for Galactic \nuc{Al}{26} at 1809\,keV.
The paper is structured as follows:
In Sect.\,\ref{sec:flight}, we summarize the COSI 2016 flight and data selections for our analysis.
The data analysis is presented in Sect.\,\ref{sec:data_analysis}.
We illustrate our results in Sect.\,\ref{sec:results}, followed by a comparison of the results with simulations in Sect.\,\ref{sec:sim_analysis}.
Finally, we discuss our results in Sect.\,\ref{sec:discussion} and summarize in Sect.\,\ref{sec:summary}.

\section{COSI}\label{sec:flight}
\subsection{The COSI 2016 Flight}\label{sec:2016flight}
On 2016 May 17, COSI was launched as a science payload on a NASA ultra-long duration balloon from Wanaka, New Zealand.
The launch site from New Zealand was chosen to maximize exposure of the Galactic Center, observations of which are important for COSI's science goals to measure nuclear lines and electron-positron annihilation.
COSI is a free-floating instrument always pointed at zenith and sweeps the sky through the Earth's rotation during flight.

A summary of the 46-day COSI 2016 flight is found in \citet{kierans2016}.
Nine of COSI's twelve detectors operated continuously throughout the flight.
Two detectors were turned off within the first 48 hours of the flight and a third was turned off on 2016 June 6.
The shut-offs were due to a well-understood high voltage problem linked to passive electronic parts which was diagnosed and fixed after the flight \citep{clio_thesis}. 
The nominal flight altitude was 33\,km, though the balloon experienced altitude variations between 33 and 22\,km with the day-night cycle.
Remaining at high altitude is preferable for balloon instruments like COSI because the strong background from Earth's albedo and atmospheric absorption decrease with increasing altitude.
Additionally, modeling the background at constant altitudes simplifies the analysis.
The instrument circumnavigated the globe within the first 14 days of the flight and then remained largely above the South Pacific Ocean before the flight was safely terminated on 2016 July 2.
The instrument was recovered from its landing site in Peru with no signs of consequential damage. 

\begin{table}
\centering
\movetableright=-0.3in
\begin{tabular}{cc}
\hline
{Region}         & {($\ell$,$b$) [$^{\circ}$]} \\
\hline
Signal           & (0 $\pm$ 30, 0 $\pm$ 10)    \\
Background Region 1 & ($-$180 $\pm$ 80, 0 $\pm$ 90) \\
Background Region 2 & (0 $\pm$ 30, 85 $\pm$ 5)    \\
Background Region 3 & (0 $\pm$ 30, $-$85 $\pm$ 5)  \\
\hline
\end{tabular}%
\caption{The longitude, latitude ($\ell$,$b$) pointing cuts defining the signal and background regions of the 2016 flight. The three background pointing cuts together comprise the background region.}
\label{table:pointing_cuts}
\end{table}

\subsection{Data selection}\label{sec:data_selection}

We select data from the 2016 flight based on previous observations of \nuc{Al}{26} and through cuts in the Compton Data Space \citep[CDS,][]{schonfelder1993instrument,zoglauer2021cosi}.
%
%\textbf{The CDS is the natural data space for Compton telescope analysis.}
%
The CDS is spanned by three parameters which specify the observed Compton scattering process as well as the measured changed state of the incident $\gamma$-ray: the Compton scattering angle ($\phi$ $\in$ [0$^{\circ}$, 180$^{\circ}$]), and the polar ($\psi$ $\in$ [0$^{\circ}$, 180$^{\circ}$]) and the azimuthal ($\chi$ $\in$ [-180$^{\circ}$, 180$^{\circ}$]) direction of the scattered $\gamma$-ray in Galactic coordinates.
These three parameters describe the arrival direction of the $\gamma$-ray. 
The event time (UTC) and photon energy of each incident photon are also recorded. 
%
%The five total parameters define the CDS. 
%
% We perform this analysis entirely within the CDS and select data using the Compton scattering angle, energy, and time. 
%\textbf{We select data inside the CDS using the Compton scattering angle, energy, and time.}
%
We integrate over the scattered $\gamma$-ray direction ($\psi$, $\chi$) since we are not performing imaging; these quantities are not relevant to the analysis described in this paper.
We use the recorded photon energy for spectral analysis and use the event time to select data from the signal and background regions of the flight.

Studies by COMPTEL and SPI show \nuc{Al}{26} emission concentrated in the Inner Galaxy ($|\ell| \leq 30^{\circ}$, $|b| \leq 10^{\circ}$), so as a conservative approach we only assume \nuc{Al}{26} emission in this well-constrained region and define the Inner Galaxy as our signal region (see Sect.\,\ref{sec:systematic_uncertainties} for further discussion about the distribution of \nuc{Al}{26} emission). 
The background region encloses the sky exclusive of the signal region.
Thus, we partition the signal and background region data by the times during which COSI's zenith pointing fell inside the respective regions.
%

% COSI's event reconstruction within the CDS constrains the origin of an incident photon to a circle on the sky defined by the Compton scattering angle $\phi$.
%COSI's event reconstruction constrains the origin of an incident photon to a circle on the sky defined by the Compton scattering angle $\phi$.
%\textbf{As an incident photon is represented in the CDS by a conical shape with an extent governed by $\phi$, the Compton scattering angle effectively broadens the field of view of the instrument.}
The Compton scattering angle effectively broadens the observation region; a zero-degree Compton scattering angle points back at the source location in image space, and an increase in the accepted Compton scattering angle will broaden this image space region by the same angle in the CDS.

\begin{figure}[!t]
\centering
\includegraphics[width=1.0\columnwidth,trim=0.5in 2.5in 0.5in 0.5in,clip=true]{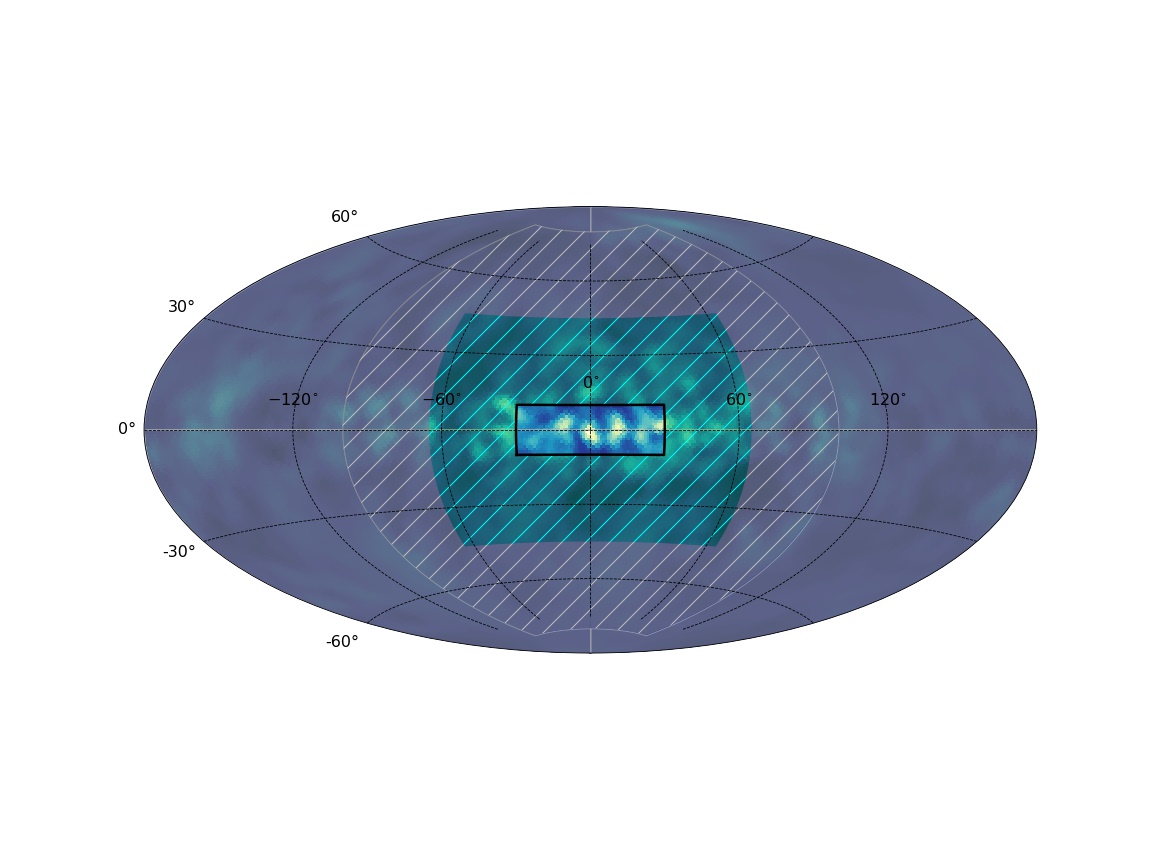}
\caption{The COSI 2016 signal and background regions (Table\,\ref{table:pointing_cuts}) displayed over the SPI \nuc{Al}{26} image \citep{Bouchet_2015}. The signal region is defined by the Inner Galaxy (black rectangular outline) and the surrounding hatched green shading maps the effective broadening of this region by the maximum Compton scattering angle $\phi_{\rm max}$ = 35$^{\circ}$. The remaining gray and hatched gray shadings map the background region and its effective 35$^{\circ}$ broadening, respectively. There is no overlap between the broadened signal and background regions. \label{fig:pointing_cuts_over_spi}}
\end{figure}

We therefore expect photons from a region extending beyond the Inner Galaxy out to a maximum Compton scattering angle $\phi_{\rm max}$ to contribute to the signal spectrum. 
To prevent overlap between the signal and background regions, the pointing cuts for the background region are chosen such that the $\phi_{\rm max}$ extensions beyond the borders of the signal and background regions fall tangential to each other (see Figure \,\ref{fig:pointing_cuts_over_spi} and Table\,\ref{table:pointing_cuts}).
We use an optimization procedure (Appendix\,\ref{appx:optimize_phi}) to define $\phi_{\rm max} = 35^{\circ}$, which yields an acceptable signal-to-noise ratio and preserves a fraction of the sky outside of the signal region large enough for sufficient background statistics.
A minimum $\phi_{\rm min} = 10^{\circ}$ removes more atmospheric background \citep{ling_model} than \nuc{Al}{26} signal events.
Thus, we apply a cut in the CDS on the Compton scattering angle $\phi$ as an optimized event selection which aims to reduce the background in the selected data. 
The signal and background regions are superimposed on the SPI 1.8\,MeV image in Figure\,\ref{fig:pointing_cuts_over_spi}.

We choose Compton events with initial energy 1750--1850\,keV and incident angle $\leq 90^{\circ}$ from COSI's zenith.
This restriction in incident angle, called the ``Earth Horizon Cut," reduces the dominant albedo background.
The number of allowed Compton scatters ranges from two to seven, the minimum distance between the first two interactions is 0.5\,cm, and that between any subsequent interactions is 0.3\,cm.
The minimum number of Compton scatters is required for reconstruction of Compton events; events with greater than seven scatters are likely to be pair-production events, which cannot be reconstructed \citep{boggs2000event}.
Imposing the minimum distances between interactions improves COSI's angular resolution.

Observations in the signal region are limited to balloon altitudes of at least 33\,km to mitigate worsening atmospheric background and attenuation with decreasing balloon altitude.
The only times disregarded in the background region are those before the balloon reached float altitude and those with high shield rates; this preserves more statistics for improved determination of the spectral shape of the background, which is not expected to change with altitude.
These event selections (Table\,\ref{table:event_selections}) result in a total observation time in the signal region of $T_{\rm {SR}} \approx 156$\,ks and that in the background region of $T_{\rm {BR}} \approx 1356$\,ks.
Given the three detector shut-offs, data from and simulations of the flight prior to 2016 June 6 are processed with a 10-detector mass model and afterwards with a 9-detector mass model. 

\begin{table*}
\centering
\begin{tabular}{cc}
\hline
{Parameter}    & {Permitted values} \\
\hline
Altitude in signal, background regions  & $\ge 33$\,km, all  \\
Energy & 1750$-$1850\,keV \\
Compton scattering angle $\phi$ & 10$^{\circ}-$35$^{\circ}$  \\
Number of Compton scatters & 2$-$7 \\
Minimum distance between the first two (any) interactions & 0.5 (0.3)\,cm \\
Earth Horizon Cut & Accept only events originating above the Earth's horizon \\
\hline
\end{tabular}%
\caption{Event selections on flight data in the signal and background regions. The resulting observation time in the signal region is 156\,ks and that in the background region is 1356\,ks.}
\label{table:event_selections}
\end{table*}

\begin{figure*}[!t]
\centering
\includegraphics[width=0.9\textwidth,trim=0.0in 0.1in 0.0in 0.1in,clip=true]{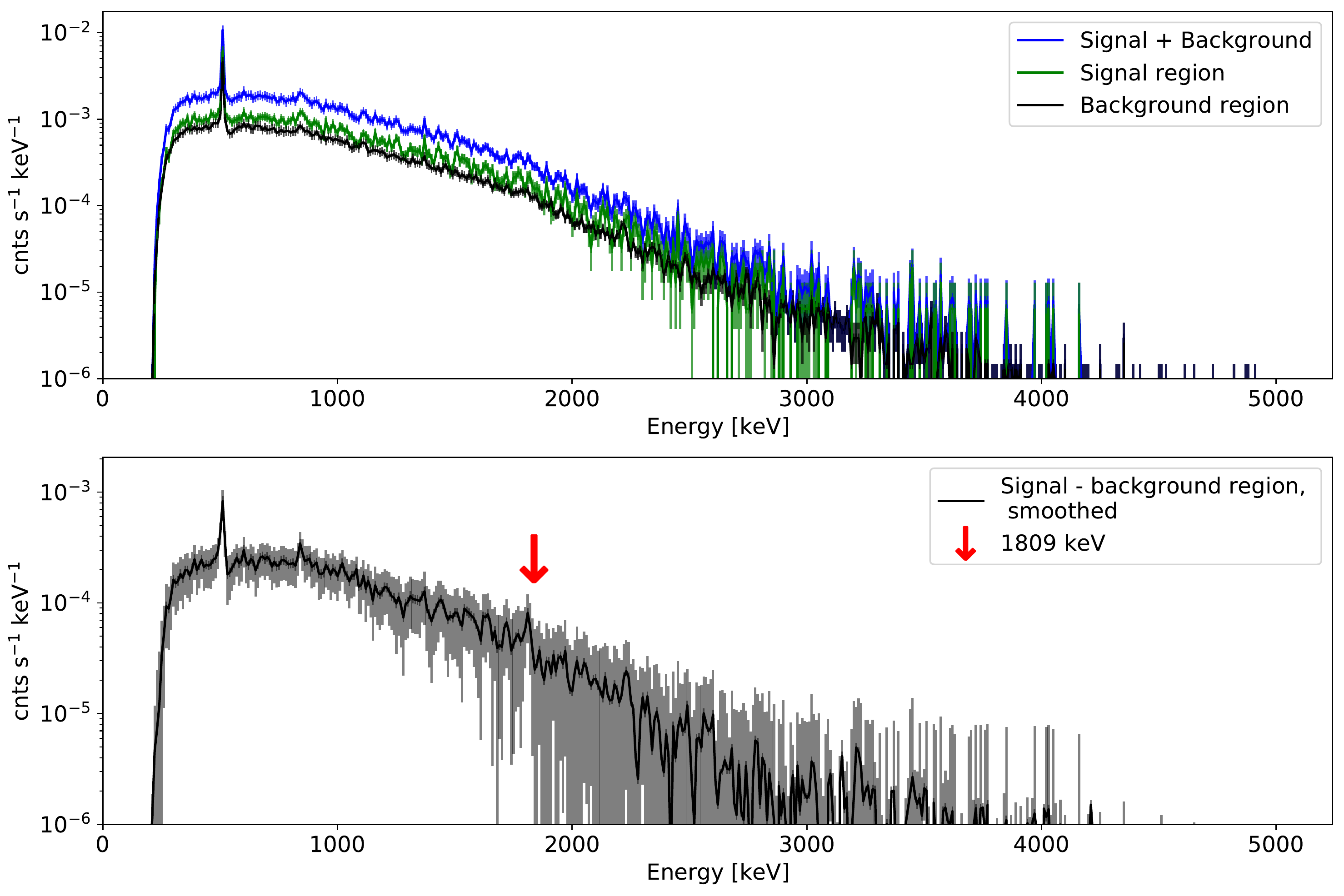}
\caption{Top: Full COSI 2016 flight spectrum of events which pass the signal and background region event selections. Bottom: Background-subtracted spectrum smoothed by a Gaussian filter of width $\sigma = 5$\,keV. Error bars are $\sqrt{\mathrm{counts}}$.}
\label{fig:full_flight_spectrum}
\end{figure*}

A full spectrum of the flight containing events which pass the signal and background region event selections is shown in Figure\,\ref{fig:full_flight_spectrum}.
The spectra are normalized by the observation time in each region.
The bottom panel is the difference of the background and signal region spectra and the result is smoothed with a Gaussian filter of width $\sigma = 5$\,keV for clarity.
In addition to the strong 511\,keV line and a general continuum, a peak near 1809\,keV is visible.

\section{Data analysis}\label{sec:data_analysis}
\begin{figure}[htp]
\includegraphics[width=1.0\columnwidth,trim=0.1in 0.0in 0.5in 0.0in,clip=true]{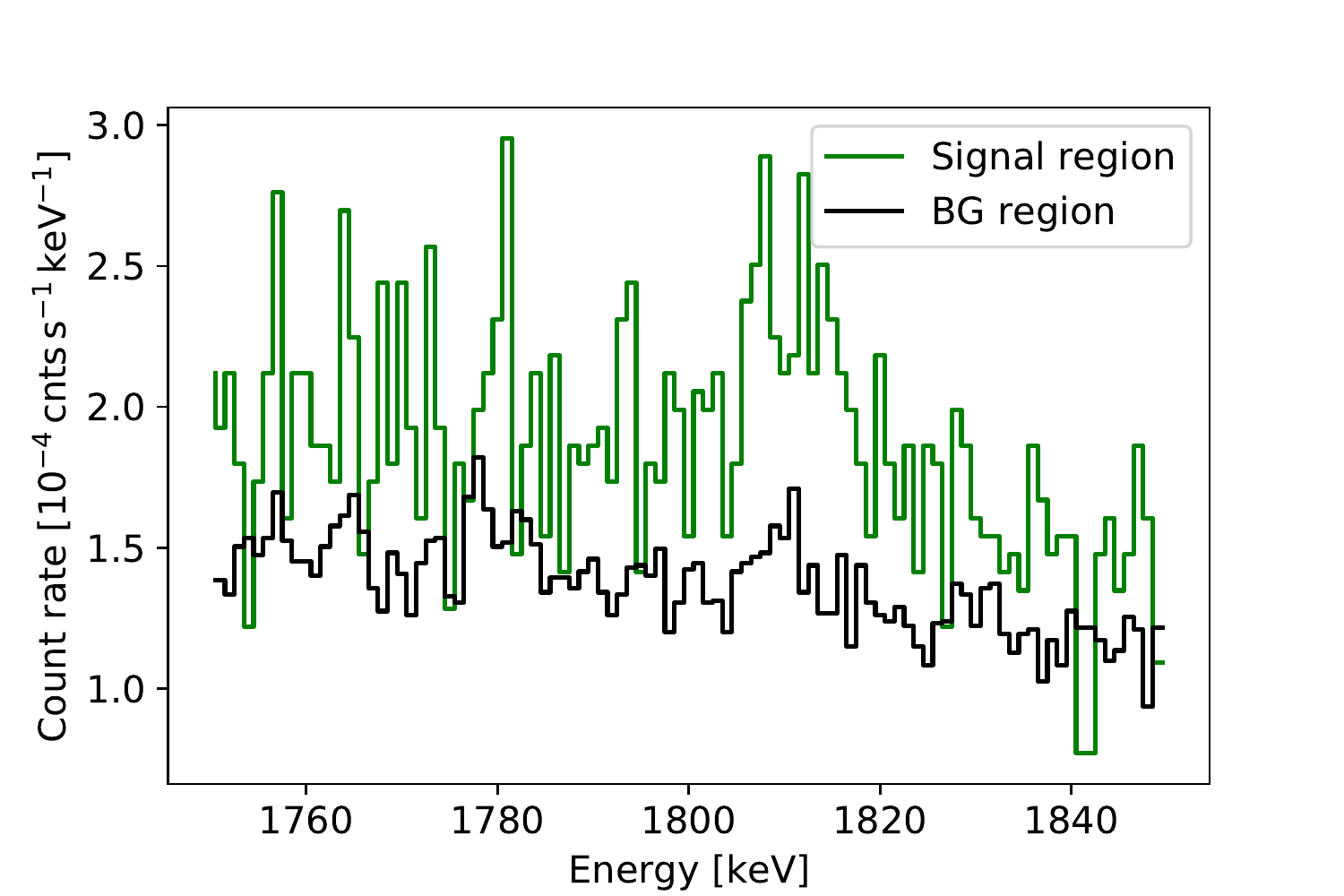}
\caption{COSI 2016 flight spectra in the signal and background regions.}
\label{fig:flight_spectra}
\end{figure}

We model COSI data, $d$, as a linear combination of a sky model, $s$, and a background model, $b$, with unknown amplitudes $\alpha$ and $\beta$, respectively.
The data are binned in 1-keV bins, $i$, spanning 1750 to 1850\,keV, such that the model reads
\begin{equation}
    m_i = \alpha s_i + \beta b_i\mathrm{.}
    \label{eq:model}
\end{equation}
\noindent The following sections describe model templates $s$ and $b$ in detail.
Photon counting is a Poisson process and the likelihood that data $d$ is produced by a model $m$ is given by the Poisson distribution
\begin{equation}
\mathcal{L}(d|m) = \prod_{i = 1}^{N} \frac{m_{i}^{d_i}e^{-m_i}}{d_{i}!}\mathrm{,} 
\label{eq:likelihood}
\end{equation}
\noindent where $N=100$ energy bins.
We fit for the scaling factors $\alpha$ and $\beta$ in the signal region data $d_i$ by minimizing the Cash statistic \citep{cash}, which is the negative logarithm of the likelihood in Eq.\,(\ref{eq:likelihood}), agnostic to model-independent terms:
\begin{equation}
\mathcal{C}(d|m) := -\sum_{i = 1}^{N} [m_{i} - d_i \mathrm{ln}(m_i)]\mathrm{.} 
\label{eq:cash}
\end{equation}
\noindent The measured data from the signal and background regions are shown in Figure\,\ref{fig:flight_spectra}. 

\subsection{Sky model}\label{sec:sky_model}
In order to construct an absolute spectral response, we simulate multiple potential realizations of the COSI 2016 measurements using the far-infrared Diffuse Infrared Background Experiment (DIRBE) $240\,\mrm{\mu m}$ map \citep{dirbe} as an image template.
We find that the expected number of photons from the signal region between 1750 and 1850\,keV is about 41.
We therefore generate 50 simulations to obtain sufficient statistics for a smooth sky model spectrum.
The flux in this bandpass is heavily dominated by \nuc{Al}{26} emission ($\sim$95\%) and we expect only a $\sim$5\% contribution from the Galactic continuum \citep{2020ApJ...889..169W}.

We use the DIRBE $240\,\mrm{\mu m}$ image because it is a good tracer of Galactic \nuc{Al}{26} emission that has been measured by COMPTEL and SPI \citep{knodlseder1999multiwavelength,Bouchet_2015}.
It also does not exhibit the weak artifacts of emission found in the SPI and COMPTEL 1.8\,MeV maps which are not easily distinguishable from true \nuc{Al}{26} emission \citep[see][]{Bouchet_2015,pluschke2001comptel}.
Furthermore, with the DIRBE $240\,\mrm{\mu m}$ image we can probe structures of emission finer than those granted by the 3$^{\circ}$ resolutions of the SPI and COMPTEL maps.
The Inner Galaxy flux of the DIRBE $240\,\mrm{\mu m}$ image is normalized to the COMPTEL \nuc{Al}{26} Inner Galaxy flux of $3.3 \times 10^{-4}$\,\flux.
The total flux in the image is $1.2 \times 10^{-3}$\,\flux.
The simulated photopeak energy is chosen as the laboratory energy of 1808.72\,keV.
Each of the 50 realizations is simulated in two parts, the first with a 10-detector mass model and the second with a 9-detector mass model, to ensure consistency with the measurements.
The transmission probability of $\gamma$-rays through the atmosphere is assumed to be constant at the selected flight altitude of 33\,km.

\begin{figure}[!t]
\includegraphics[width=1.0\columnwidth]{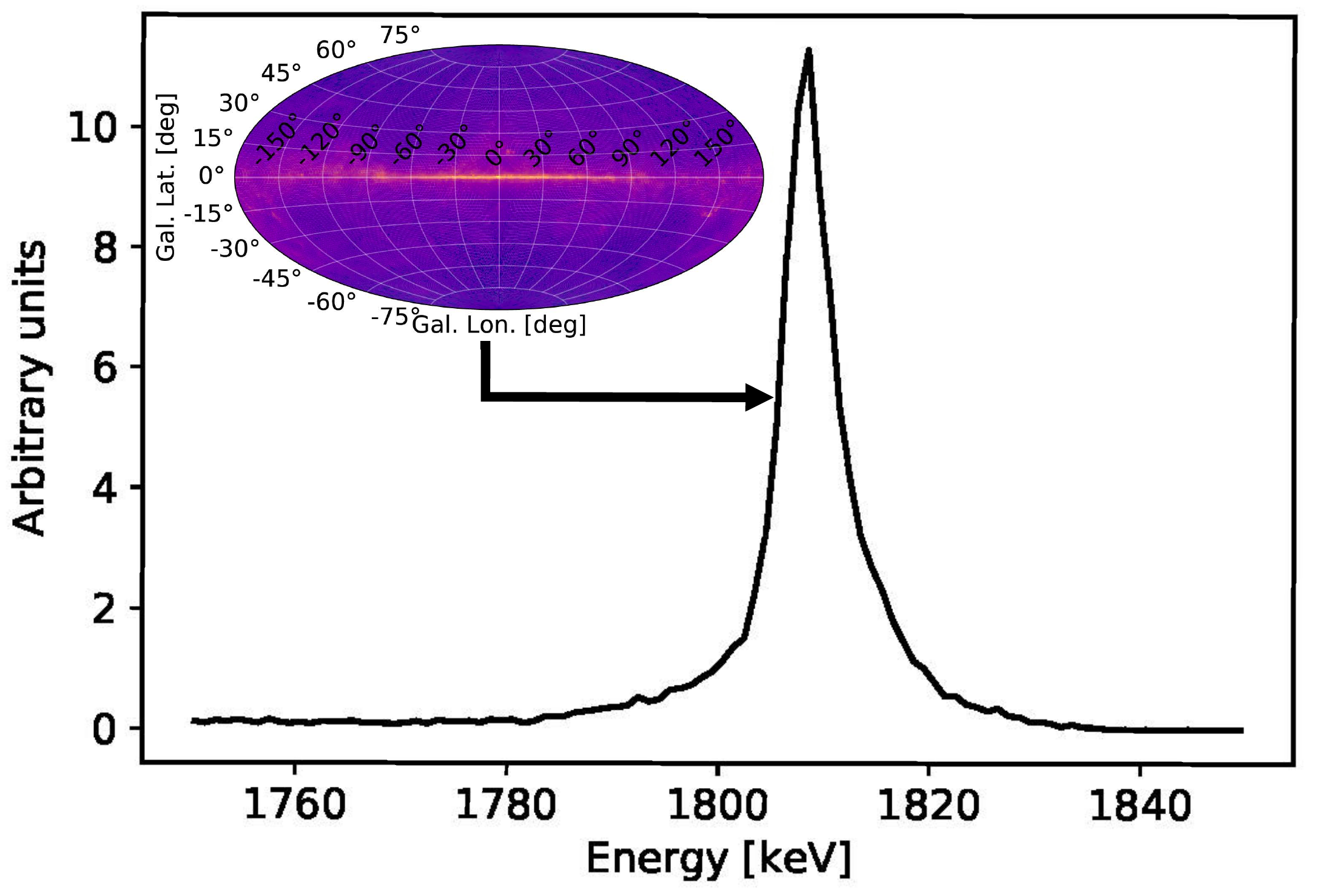}
\caption{The spectral sky model defined by COSI's response to the DIRBE $240\,\mrm{\mu m}$ map (inset image) over 50 2016 flights.}
\label{fig:sky_model}
\end{figure}

Figure\,\ref{fig:sky_model} shows the energy spectrum of events simulated over 50 realizations of the DIRBE $240\,\mrm{\mu m}$ map which pass the event selections described in Sect.\,\ref{sec:data_selection}.
This spectrum defines the sky model.
The tailing above 1809\,keV is possibly a consequence of increased cross-talk between strips at high energies, which artificially enhances the recorded energy of an event, and complications in event reconstruction at high energies.
In particular, 17\% of events at 511\,keV and 28\% of events at 1275\,keV cannot be accurately reconstructed \citep{clio_thesis}.
Applying this same reconstruction check to \nuc{Al}{26} all-sky simulation reveals that $\sim$ 30\% of events at 1809\,keV are too complicated to reconstruct.
However, this complication does not prohibit \nuc{Al}{26} analysis of real flight data because COSI’s complete spectral response is generated using the same reconstruction algorithm. 
The complication is thus represented in the sky model and simulations.

\subsection{Background model}\label{sec:bg_model}
As a data-driven approach to background modeling which draws upon the expectation that \nuc{Al}{26} emission is concentrated in the Inner Galaxy, we infer a background model from high latitudes.
Recent discussion in the literature about high-latitude emission of \nuc{Al}{26} \citep{pleintinger2019comparing,rodgers-lee} competes with this assumption of concentrated Inner Galactic emission. 
However, high-latitude emission of \nuc{Al}{26} remains unconstrained against the well-established emission from the Inner Galaxy.
Additionally, if the high-latitude emission is of extragalactic origin, then it will also be present behind the Inner Galaxy. 
In that case it is necessary to account for it as background in a measurement of the Inner Galaxy.
Thus, we proceed with our expectation of dominant Inner Galactic emission.
Regions outside the Inner Galaxy remain valid contributors to our estimation of the background spectrum.
Systematic uncertainties from this assumption are discussed in Sect.\,\ref{sec:systematic_uncertainties}.

We probe the underlying shape of the background spectrum in Figure\,\ref{fig:flight_spectra} with an empirical fit to data in the background region.
For enhanced statistics, these data are considered with minimal event selections compared to those outlined in Sect.\,\ref{sec:data_selection}, limited only to Compton events of incident energy 1750--1850\,keV and Compton scattering angles $\phi$ $\leq 90^{\circ}$.
We use a power law plus $N_{\ell} = 3$ Gaussian-shaped lines to provide a smooth description of and evaluate uncertainties in the measured background:
\begin{equation}
\small
    b(E) = C_0 \left( \frac{E}{E_c} \right)^{\gamma} + \sum_{l = 1}^{3} \frac{A_l}{\sqrt{2\pi}\sigma_l} \exp\left(-\frac{1}{2} \left( \frac{E-E_l}{\sigma_l} \right)^2 \right)\mathrm{.}
    \label{eq:bg_model}
\end{equation}
\noindent The first term of Eq.\,(\ref{eq:bg_model}) describes the continuum emission from atmospheric background with a power law of amplitude $C_0$, pivotal energy $E_c = 1.8$\,MeV, and index $\gamma$.
The three Gaussian-shaped lines $\ell$ are parameterized by their rates $A_l$, centroids $E_l$, and widths $\sigma_l$.

The fit of Eq.\,(\ref{eq:bg_model}) to the background spectrum is shown in Figure\,\ref{fig:bg_model} and the fitted parameters are listed in Table\,\ref{table:flight_data_bg_priors} of Appendix \ref{sec:appendix_figures_tables}.
The Gaussian-shaped lines are due to excitation of materials in the instrument payload which decay on the timescale of the flight. 
The exact origins of these instrumental lines are uncertain but appear in various other experiments with similar instrument materials \citep{mahoney1984heao,malet1991observation,naya1997gris,ayre1984line,boggs2000event,spi_bg_lines}.
The line near 1764\,keV is commonly identified as the decay of natural $\mathrm{^{238}U}$.
The 1779\,keV line is likely from the neutron capture process $\mathrm{^{27}Al}(n,\gamma)\mathrm{^{28}Al}$ followed by the 1779\,keV $\gamma$-ray emission from $\mathrm{^{28}Al}(\beta^{-})\mathrm{^{28}Si}$.
The line near 1808\,keV is likely a blend of activation lines, for example $\mathrm{^{27}Al}(n,np)\mathrm{^{26}Mg^{*}}$ and \nuc{Na}{26}($\beta^{-}$)$\mathrm{^{26}Mg^{*}}$ which then de-excite to $\mathrm{^{26}Mg}$.
The decay of \nuc{Mn}{56}($\beta^{-}$)$\mathrm{^{56}Fe^{*}}$, which produces a line at 1810.9\,keV of similar intensity to the signal 1808.7\,keV line in the background spectrum of SPI \citep{spi_bg_lines}, could also contribute to the blend.
The empirical approach to modeling the background attempts to capture these lines, whose centroids differ by less than the instrumental energy resolution. 
The spectral shapes and uncertainties of the fit shown in Figure\,\ref{fig:bg_model} are then included as normal priors to the simultaneous fit of the background and signal regions, discussed in the next section.

\begin{figure}[htpb]
\includegraphics[width=1.0\columnwidth,trim=0.3in 0.1in 0.8in 0.4in,clip=true]{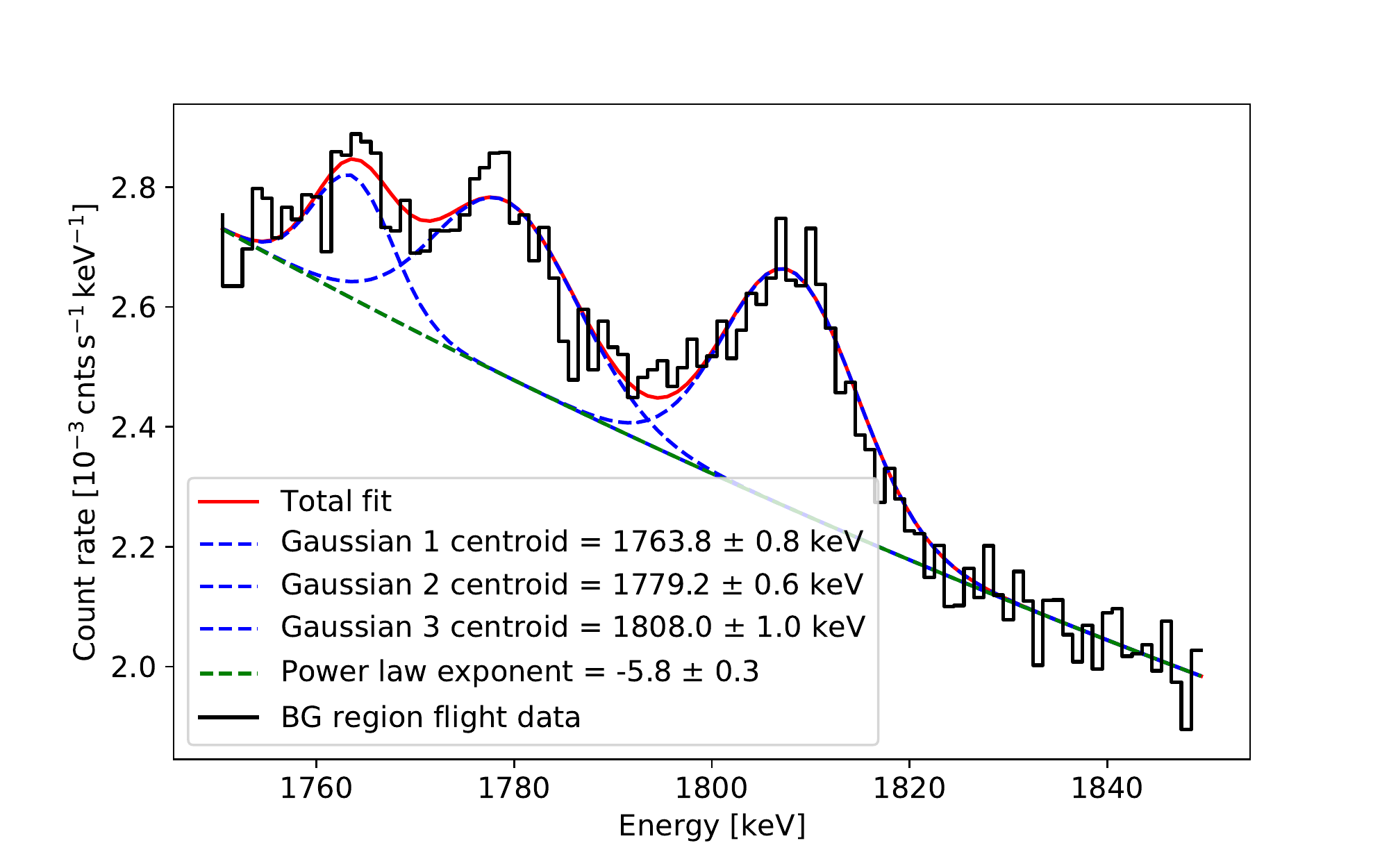}
\caption{Empirical fit to COSI flight data in the background region, with minimal event selections, which provides a smooth description of the background template shape. The fitted parameters are listed in Table\,\ref{table:flight_data_bg_priors} of Appendix \ref{sec:appendix_figures_tables}.}
\label{fig:bg_model}
\end{figure}

\begin{figure}[htpb]
\centering
\includegraphics[width=1.0\columnwidth,trim=0.0in 0.4in 0.2in 0.3in,clip=true]{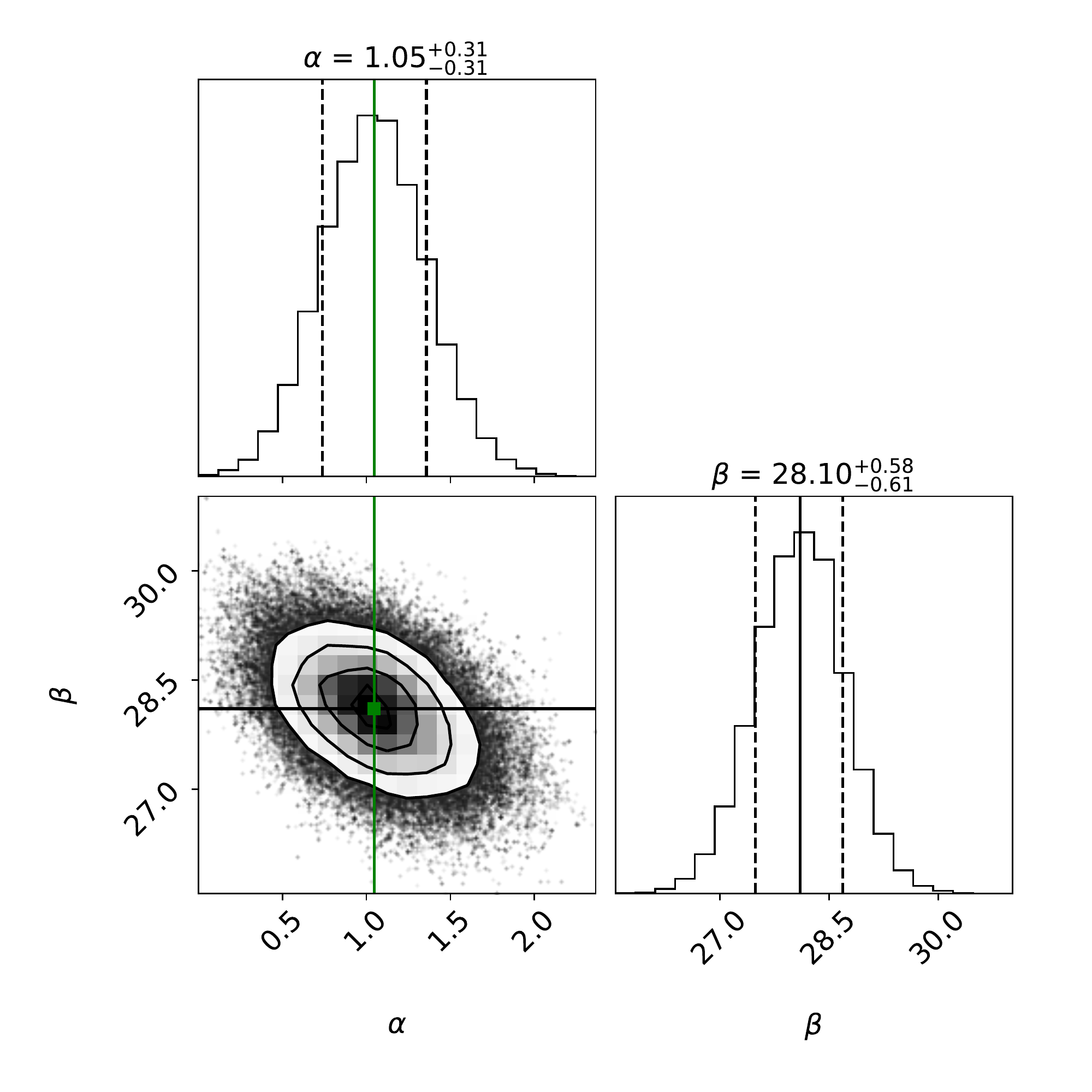}
\caption{Posterior distributions of the sky amplitude $\alpha$ and background amplitude $\beta$ in the COSI 2016 signal region. The green and black lines indicate the median $\alpha$ and $\beta$, respectively.}
\label{fig:flight_data_alpha_beta}
\end{figure}

\subsection{Propagating background uncertainties in a joint fit}\label{sec:flight_data_joint_fit}
We mitigate the potential for bias introduced by the noisy background spectrum in Figure\,\ref{fig:flight_spectra} by including the spectral features of the fit to the minimally-constrained background spectrum (Figure\,\ref{fig:bg_model}) in a subsequent, simultaneous fit of the sky and background models.
We do not expect the spectral shape of the background to vary significantly during the 46-day flight and allow the complete background model $b(E)$ to vary only within the uncertainties of the parameters from the background region fit (Sect.\,\ref{sec:bg_model}).
The continuum slope and amplitude are left variable to account for possible continuum emission in the signal region.
Therefore, this procedure only detects $\gamma$-ray lines and suppresses any instrumental as well as celestial continuum contribution.
We note that the extended Galactic Plane continuum emission from Inverse Compton scattering might readily be visible with COSI (see continuum emission in Figure\,\ref{fig:full_flight_spectrum}) in a separate analysis which does not suppress the continuum as background.
Thus, by using Eq.\,(\ref{eq:model}), we optimize for $\alpha$ and $\beta$ accounting for the 11 known but uncertain background parameters.
The only constraint (prior) for $\alpha$ and $\beta$ is to be positive definite.
The likelihood, Eq.\,(\ref{eq:likelihood}), is therefore used to construct a joint posterior distribution by including the uncertainties in Table\,\ref{table:flight_data_bg_priors} as normal priors.
We use \texttt{emcee} \citep{emcee} to estimate the posterior distribution by Monte Carlo sampling.
The final fit values of the continuum are $C_0$ = (1.13 $\pm$ 0.02) $\times$ 10$^{-3}$ cnts s$^{-1}$ keV$^{-1}$ and $\gamma = -$4.1 $\pm$ 0.6.
This is considerably different from the background-only region, suggesting that the celestial continuum is absorbed in the background model fit and that COSI can readily measure the extended Galactic Plane continuum. The latter is beyond the scope of this paper.

As a check of consistency, we compare the amplitudes of the three Gaussian-shaped lines in the empirical fit to the background region data (Figure\,\ref{fig:bg_model}, Table\,\ref{table:flight_data_bg_priors}) and the amplitudes returned by this simultaneous fit to the signal region data in Figure\,\ref{fig:flight_spectra}.
We call the $\sim$1764\,keV, $\sim$1779\,keV, and $\sim$1808\,keV peak amplitudes A1, A2, and A3, respectively, per the notation in Table\,\ref{table:flight_data_bg_priors}.
Normalizing all amplitudes to A1, we find amplitude ratios in the empirical background fit of A1/A1 $\sim 1.0 \pm 0.4$, A2/A1 $\sim 2.6 \pm 0.4$, and A3/A1 $\sim 3.3 \pm 0.3$. 
Those in the simultaneous fit are A1/A1 $\sim 1.0 \pm 0.4$, A2/A1 $\sim 2.2 \pm 0.5$, and A3/A1 $\sim 2.4 \pm 0.5$.
The ratios are consistent within 1$\sigma$ uncertainties.

\section{Results}\label{sec:results}
\subsection{Signal region}\label{sec:signal_region}
We find an expected dominance of background with best-fit values of $\alpha = 1.1 \pm 0.3$ and $\beta = 28.1 \pm 0.6$ (Figure\,\ref{fig:flight_data_alpha_beta}).
Amplitudes $\alpha$ and $\beta$ represent the number of photons per keV emitted by the sky and background, respectively.
An $\alpha$ value consistent with zero would imply that the signal region data are entirely explained by the background model only.
Hence, from $\alpha$ we derive a signal-to-noise ratio, as estimated by the best-fit amplitude compared to its uncertainty, of $1.1/0.3 \sim 3.7$.

A maximum likelihood ratio calculation \citep{li1983analysis} formalizes the significance of the measurement above background.
This ratio $\lambda$ is defined as 
\begin{equation}
    \lambda = \ln L(D|\alpha,\beta) - \ln L(D|\alpha=0,\beta),
    \label{eq:max_likelihood_ratio}
\end{equation}
\noindent where $L(D|\alpha,\beta)$ is the likelihood of the simultaneous fit including non-zero sky and background model contributions. The second term, $L(D|\alpha=0,\beta)$, is the likelihood that the signal region data are explained solely by the background (the null hypothesis).
The significance $\sigma$ of the measurement above background is then calculated as the square-root of the test statistic $\mrm{TS} = 2\lambda$, such that
\begin{equation}
    \sigma = \sqrt{TS} = \sqrt{2\lambda}\mathrm{.}
    \label{eq:TS_significance}
\end{equation}
\noindent This calculation yields a 3.7$\sigma$ significance above background of the 1809\,keV \nuc{Al}{26} peak in COSI 2016 flight data.
Multiplying the measured rate of $6.8 \times 10^{-4}\,\mrm{cnts\,s^{-1}}$ between 1750 and 1850\,keV by the exposure time $T_{\rm {SR}}$ gives $\sim$106 \nuc{Al}{26} photons.
%
%The background rate of $2.6 \times 10^{-3}\,\mrm{cnts\,s^{-1}}$ between 1803 and 1817\,keV gives $\sim$407 background photons.
The background rate of $3.0 \times 10^{-4}\,\mrm{cnts\,s^{-1}}$ between 1803 and 1817\,keV gives $\sim$407 background photons.

\begin{figure}[htpb]
\includegraphics[width=1.0\columnwidth,trim=0.0in 0.0in 0.7in 0.6in, clip=true]{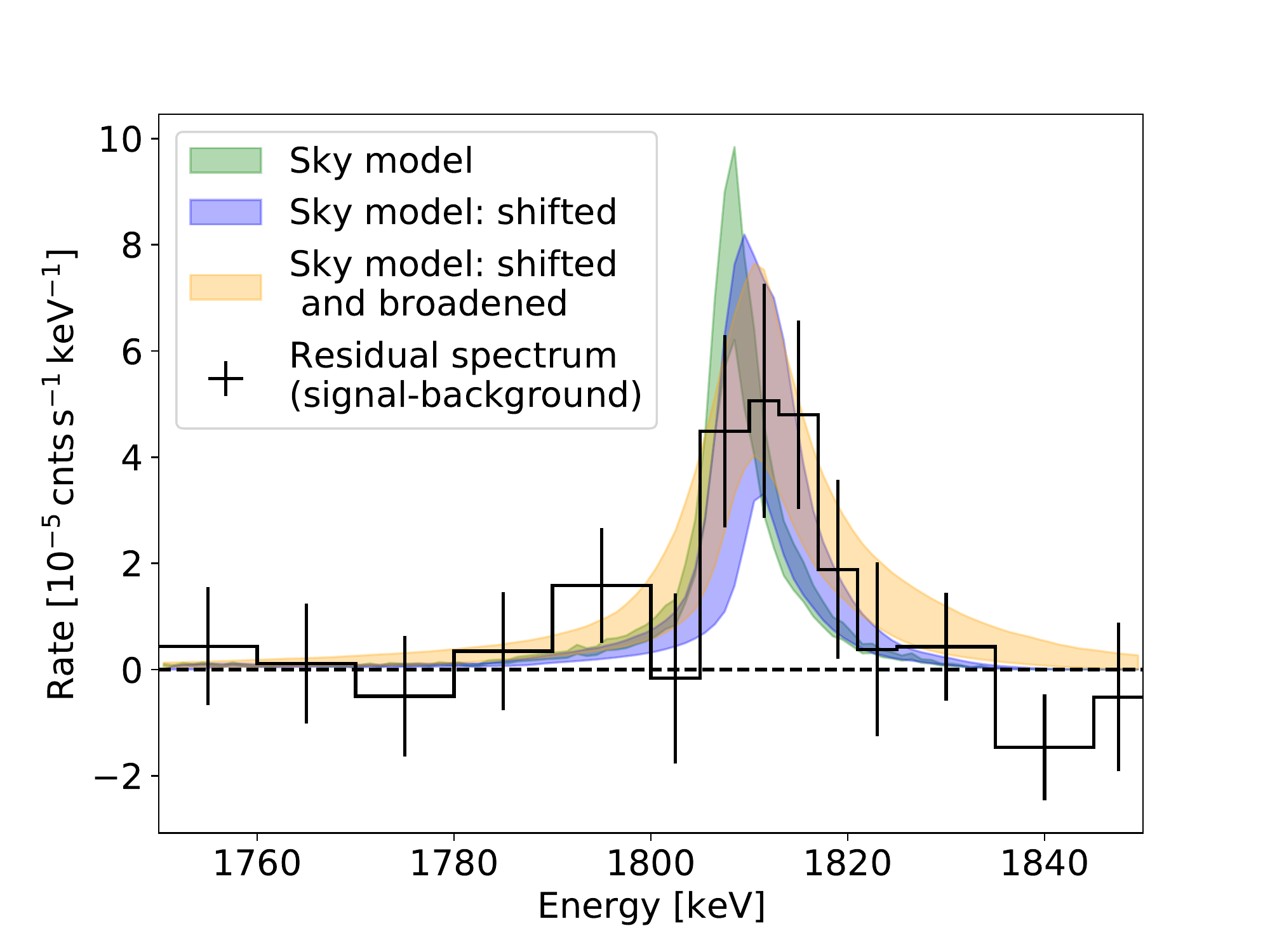}
\caption{Background-subtracted spectrum from the COSI 2016 flight. The 1$\sigma$ contours of the sky models when we fit for line shift ($\Delta E$ = 2.5 $\pm$ 1.8\,keV) and combined shift and broadening ($\Delta E$ = 2.9 $\pm$ 1.4\,keV, intrinsic sky broadening $<$ 9.7\,keV (2$\sigma$ upper limit)) are also shown.}
\label{fig:flight_data_bg_sub_spec}
\end{figure}

The background-subtracted spectrum is provided in Figure\,\ref{fig:flight_data_bg_sub_spec}.
Note that the count rates near the prominent background lines at 1764 and 1779\,keV (Figure\,\ref{fig:bg_model}) are consistent with zero.
This is validation of our background handling method.

\subsection{Line parameters} \label{sec:line_parameters}
A summary of line parameters from the COSI 2016 flight is provided in Table \ref{table:flight_data_summary}. We use the ratio of fitted \nuc{Al}{26} counts in the signal region to the number of \nuc{Al}{26} counts expected from DIRBE $240\,\mrm{\mu m}$ all-sky simulations to calculate COSI's measured \nuc{Al}{26} flux.
The ratio between the fitted flight and simulated counts is $\sim 2.6$. 

\begin{figure}[htpb]
\includegraphics[width=1.0\columnwidth,trim=0.0in 0.0in 0.5in 0.4in, clip=true]{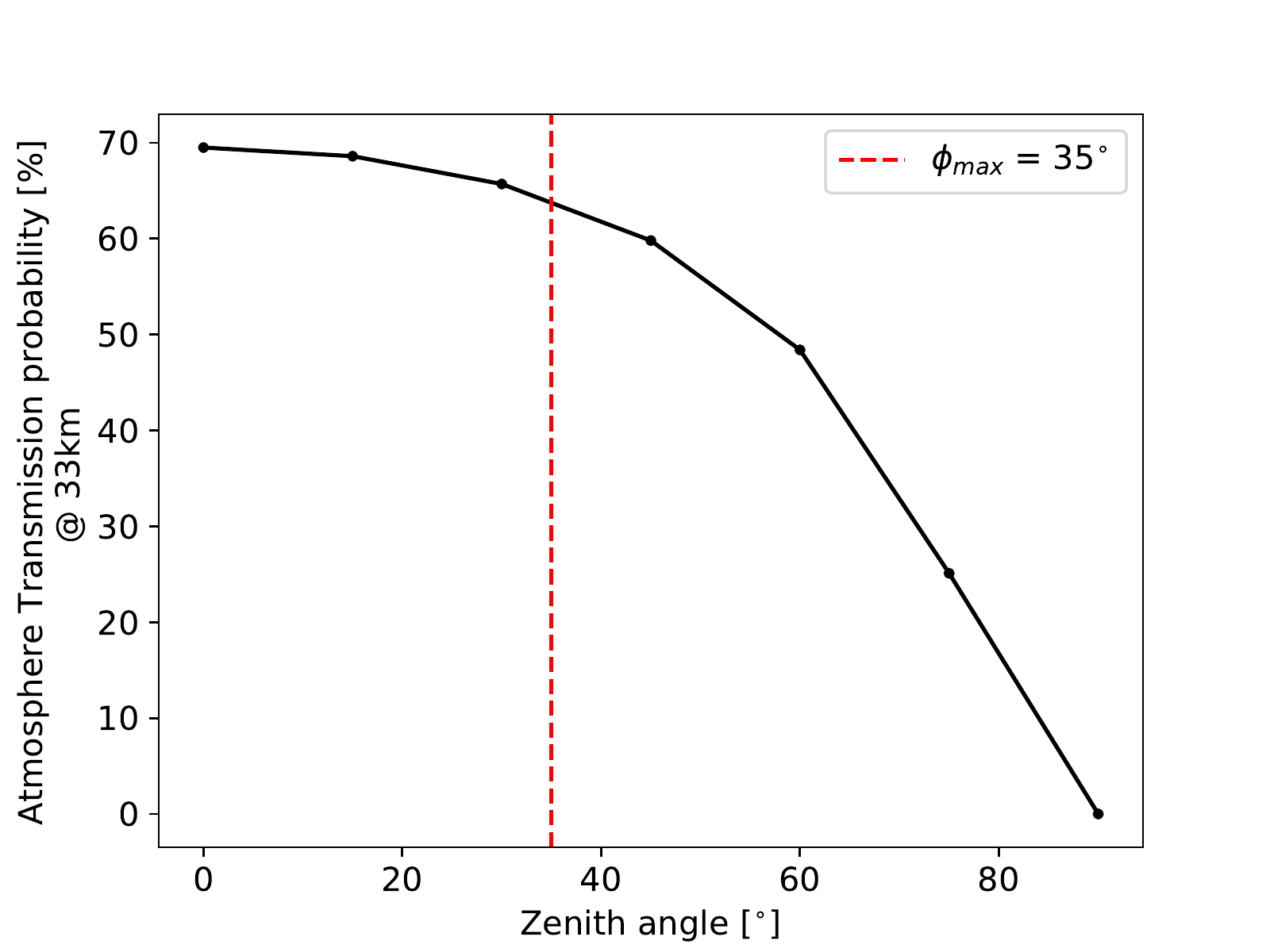}
\caption{Zenith response of COSI to 2\,MeV photons at a flight altitude of 33\,km, indicating strongest sensitivity to photons originating from within $\leq 35$--$40^{\circ}$.}
\label{fig:zenith_resp}
\end{figure}

Using atmospheric transmission data from NRLMSISE-00 \citep{nrlmsise-00}, we find that the response of COSI near 1.8\,MeV at 33\,km altitude exhibits a sharp decrease in the number of photons beyond a zenith angle of 35--40$^{\circ}$ (Figure\,\ref{fig:zenith_resp}).
As such, we expect COSI to be sensitive to photons out to $\sim$35$^{\circ}$ beyond the specified Inner Galaxy pointing cut. 
We also defined the maximum Compton scattering angle as 35$^{\circ}$ (Appendix\,\ref{appx:optimize_phi}).
Assuming that the true flux follows the DIRBE $240\,\mrm{\mu m}$ image, we report a measured COSI 2016 \nuc{Al}{26} flux of $(1.70 \pm 0.49) \times 10^{-3}$\,\flux in this broadened region $|\ell|$ $\leq$ 65$^{\circ}$, $|b|$ $\leq$ 45$^{\circ}$.
The COSI 2016 measurement of flux from the Inner Galaxy ($|\ell|$ $\leq$ 30$^{\circ}$, $|b|$ $\leq$ 10$^{\circ}$) is $(8.6 \pm 2.5) \times 10^{-4}$\,\flux.
%

% Using the COMPTEL 1.8\,MeV image as a template map instead of the DIRBE $240\,\mrm{\mu m}$ image yields an Inner Galaxy flux of $(6.6 \pm 1.9) \times 10^{-4}$\,\flux. 
% %
% Using the SPI 1.8\,MeV image gives $(7.3 \pm 2.1) \times 10^{-4}$\,\flux. 
% %
% The COSI 2016 Inner Galaxy flux values across template maps are therefore consistent with each other within uncertainties. 
% %

Next, we fit for a shift in the line centroid from the \nuc{Al}{26} laboratory energy of 1808.72\,keV to probe dynamics of the emission.
\cite{Kretschmer_2013} measure a maximum shift of $\sim$ 300\,\vel, corresponding to $\sim$1.8\,keV at 1809\,keV. 
Including systematic uncertainties from instrument calibrations, the line shift could be at most 3\,keV, or $\sim$500\,\vel.
To estimate the line centroid in the flight data, we assume that the spectral response within our 1750--1850\,keV energy window is constant.
We use a spline interpolation of the sky model template and invoke a scale parameter $\Delta E$ that shifts the total spectrum along the energy axis.
Since at small velocities the Doppler shift is proportional to the difference in centroid energy, $\Delta E$ provides a direct measure of the line shift.
By including $\Delta E$ as a free parameter in our model, we find a shift of $\Delta E = 2.5 \pm 1.8$\,keV for a centroid energy of $E_{\rm sky} = 1811.2 \pm 1.8$\,keV, and a line flux in the Inner Galaxy of $(8.8 \pm 2.5) \times 10^{-4}$\,\flux.
The 1$\sigma$ contour of this shifted sky model is plotted over the background-subtracted spectrum in Figure\,\ref{fig:flight_data_bg_sub_spec}.

\begin{figure}[htpb]
\includegraphics[width=1.0\columnwidth,trim=0.0in 0.0in 0.0in 0.0in, clip=true]{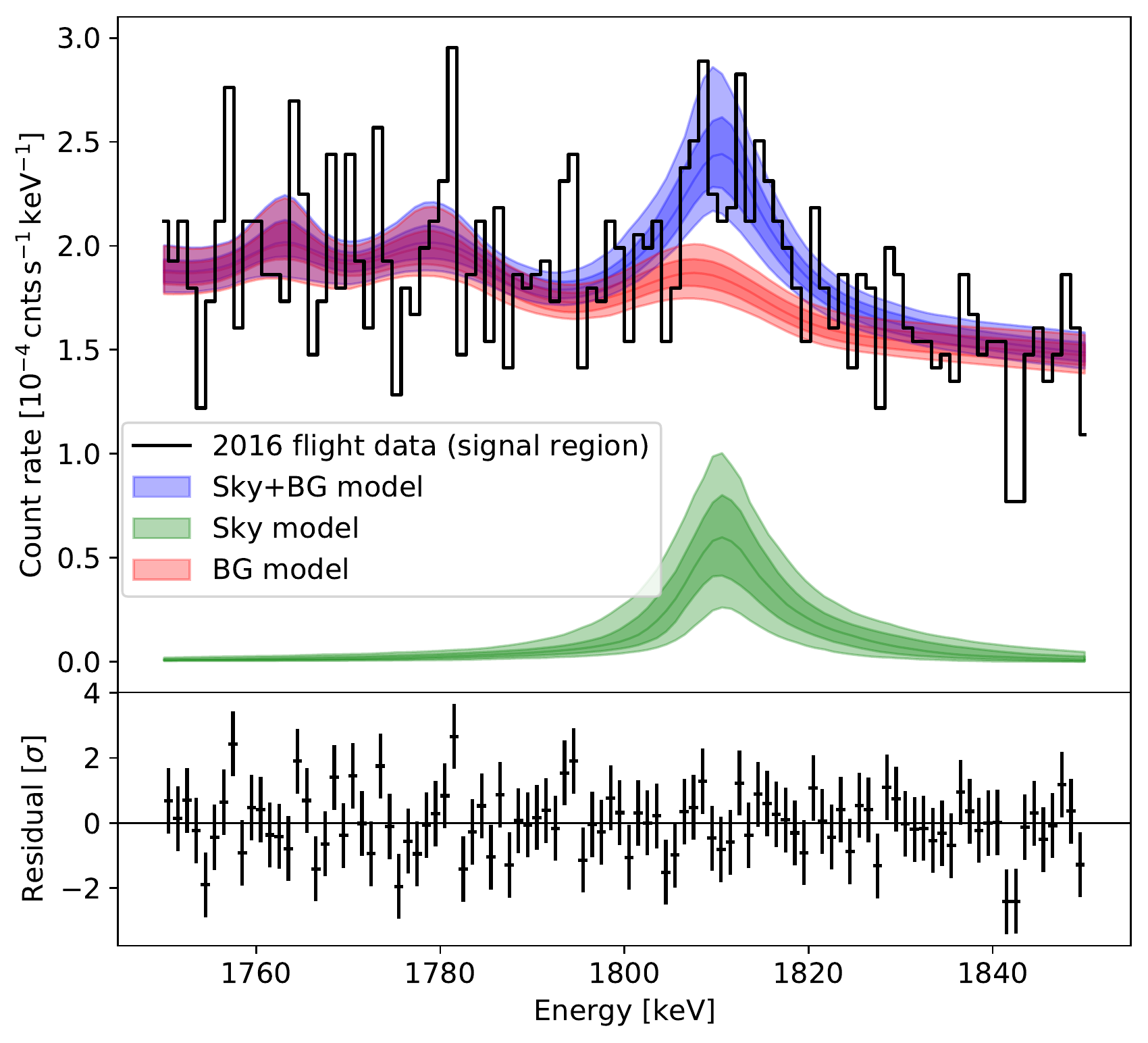}
\caption{Top: Summed (Sky+BG) and individual sky and background models plotted over the flight signal region spectrum. The sky model shown here includes the fitted energy shift and broadening parameters. The medians of the models are shown as solid lines with their 1$\sigma$ and 2$\sigma$ uncertainties as shaded contours. Bottom: Normalized residuals of the fit.}
\label{fig:flight_data_total_models}
\end{figure}

We also include a free parameter to estimate the broadening of the line. Fitting for both the line shift and broadening, we obtain a shift of $\Delta E = 2.9 \pm 1.4$\,keV and a 2$\sigma$ upper limit on the intrinsic sky broadening of 9.7\,keV. %
The 2$\sigma$ upper limit on the turbulent velocity of the \nuc{Al}{26} ejecta is $\sim 2800$\,km s$^{-1}$. 
The fit of the total model to the data, with the shifted and broadened sky model, is shown in Figure\,\ref{fig:flight_data_total_models}.
The 1$\sigma$ contour of this shifted and broadened sky model is also shown in Figure\,\ref{fig:flight_data_bg_sub_spec} and the line flux is enhanced by $\sim 30$\%.

\begin{table}
\centering
\begin{tabular}{cc}
\hline
{Line parameter}  & {Value}  \\
\hline
Measurement significance & 3.7$\sigma$ \\
Inner Galaxy flux  & (8.6 $\pm$ 2.5) $\times$ 10$^{-4}$\,\flux \\
Centroid & 1811.2 $\pm$ 1.8 keV \\
Intrinsic sky broadening (2$\sigma$) & $<$ 9.7 keV \\
Turbulent velocity (2$\sigma$) & $<$ 2800 \vel  \\
\hline
\end{tabular}%
\caption{\nuc{Al}{26} line parameters from the COSI 2016 flight. The chosen template map is the DIRBE $240\,\mrm{\mu m}$ image and the quoted uncertainties are statistical.}
\label{table:flight_data_summary}
\end{table}

\subsection{Method validation}
\label{sec:flight_method_validation}
We repeat the flight data analysis under a variety of assumptions in order to validate the method and define systematic uncertainties (Sect.\,\ref{sec:flight_data_systematics}).
Sect.\,\ref{sec:diff_template_maps} tests the method with the COMPTEL 1.8\,MeV and SPI 1.8\,MeV images as template maps. 
The subsequent tests use the DIRBE $240\,\mrm{\mu m}$ image.

\subsubsection{Different template maps}
\label{sec:diff_template_maps}
Using the COMPTEL 1.8\,MeV image as a template map instead of the DIRBE $240\,\mrm{\mu m}$ image yields an Inner Galaxy flux of $(6.6 \pm 1.9) \times 10^{-4}$\,\flux with 3.6$\sigma$ significance. 
Using the SPI 1.8\,MeV image gives $(7.3 \pm 2.1) \times 10^{-4}$\,\flux with 3.7$\sigma$ significance. 
The COSI 2016 Inner Galaxy flux values across template maps are therefore consistent with each other within uncertainties.

\subsubsection{Signal region altitude}
\label{sec:method_validation_sig_altitude}
\begin{figure}[htpb]
\includegraphics[width=1.0\columnwidth,trim=0.1in 0.0in 0.0in 0.4in, clip=true]{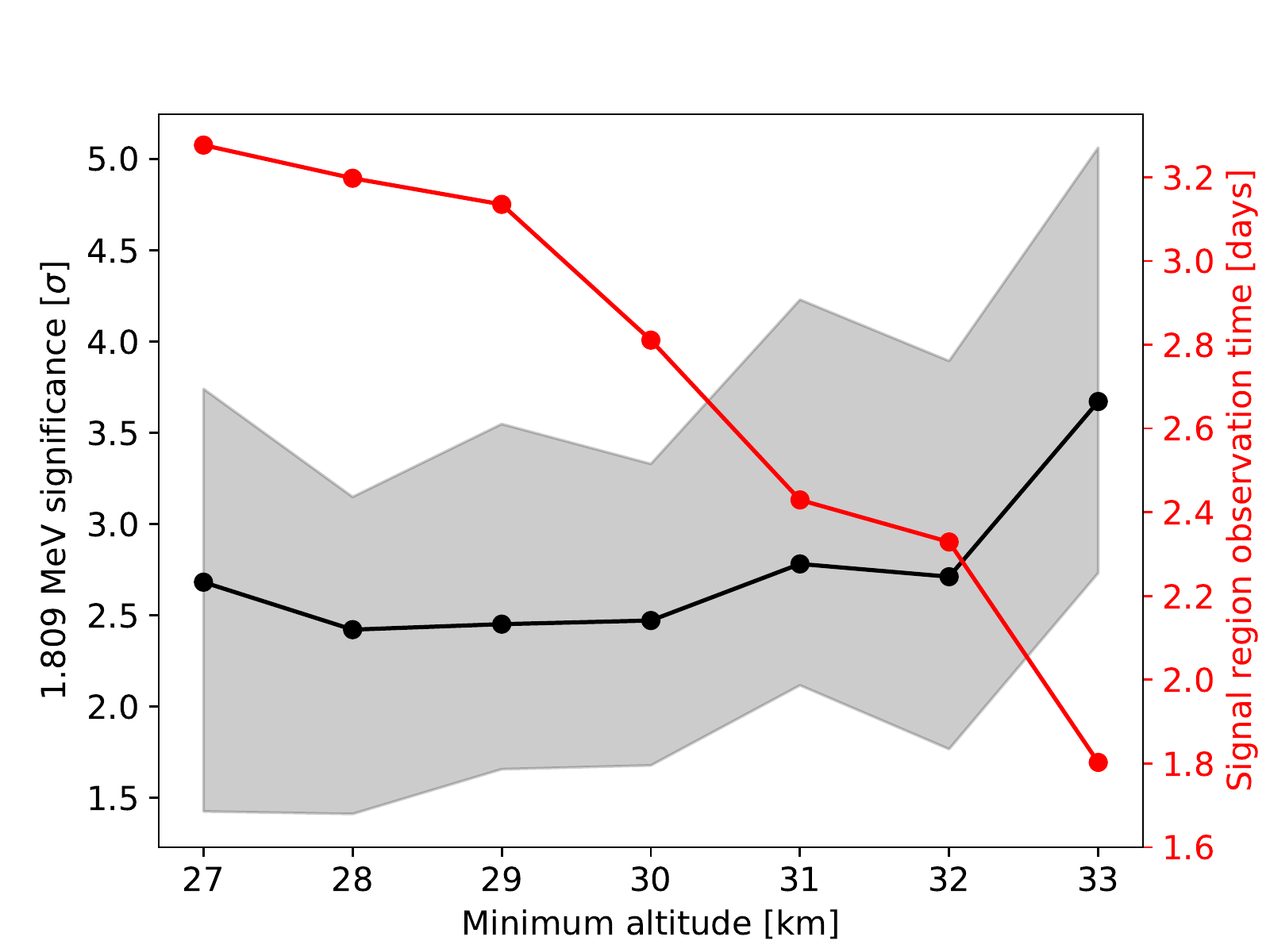}
\caption{Significance above background of the \nuc{Al}{26} measurement  as a function of minimum flight altitude. Black points: significance from flight data. Gray contour: $1\sigma$ uncertainties from 25 Poisson samples of the flight data signal region spectrum. Red points: signal region observation time from flight data.}
\label{fig:sig_vs_altitude}
\end{figure}

As a check on the consistency of our maximum-likelihood framework, we repeat the analysis considering flight data in the signal region from decreasing minimum altitudes.
We observe an expected decrease in measurement significance as atmospheric background and absorption increase (black points in Figure\,\ref{fig:sig_vs_altitude}).
%
%To estimate a spread in the significance, we analyze 25 Poisson samples of the signal region spectrum from the flight data at each altitude. 
%
To estimate a spread in the significance, we generate simulated data sets by drawing 25 Poisson samples from the signal region flight spectrum at each altitude. 
These simulated realizations of the real data contain different numbers of photons, resulting in significance values with some scatter.
The mean and standard deviation of these 25 scattered significance values per altitude define the gray $1\sigma$ contour in Figure\,\ref{fig:sig_vs_altitude}.
The severity of background contamination at balloon altitudes is especially clear, given that the observation time gained by permitting lower altitude observations cannot compensate for the worsening background environment.

We also record the Inner Galaxy flux for each minimum altitude, corresponding to each black point in Figure\,\ref{fig:sig_vs_altitude}. 
The minimum flux is (6.8 $\pm$ 2.9) $\times$ 10$^{-4}$\,\flux at a minimum altitude of 30\,km and the maximum is the (8.6 $\pm$ 2.5) $\times$ 10$^{-4}$\,\flux measurement at a minimum altitude of 33\,km in the signal region.
The flux values therefore range from (3.9--11.1) $\times$ 10$^{-4}$\,\flux.

\subsubsection{Background region altitude}
To conform with the event selections of the signal region, we apply a 33\,km minimum altitude cut in the background region and repeat the analysis. 
We measure \nuc{Al}{26} with 3.6$\sigma$ significance above background and find an Inner Galaxy flux of (8.3 $\pm$ 2.5) $\times$ 10$^{-4}$\,\flux.
This is consistent with the originally presented results. %

\subsubsection{Separate 10-, 9-detector portions}
We separate the data from the first half (10-detector portion) and second half (9-detector portion) of the flight and repeat the analysis procedure on each subset.
Using only 10-detector data, we measure \nuc{Al}{26} with 2.3$\sigma$ significance above background and find an Inner Galaxy flux of (6.8 $\pm$ 3.0) $\times$ 10$^{-4}$\,\flux.
Using only 9-detector data, we find 2.0$\sigma$ significance above background and an Inner Galaxy flux of (8.1 $\pm$ 4.1) $\times$ 10$^{-4}$\,\flux.
Within uncertainties, these results are consistent with those of the combined data set.
The significance of the measurement in the first part of the flight is slightly greater than that in the second part of the flight because COSI had more exposure to the signal region in the former.
Thus, despite the lower background during the second part of the flight, we see a stronger signal in the higher background conditions of the first half. 
Combining the data from both parts of the flight gives the strongest signal.

\subsubsection{Rigidity}
\label{sec:rigidity}
In Figures\,\ref{fig:full_flight_spectrum} and \ref{fig:flight_spectra}, we are agnostic to changes in geomagnetic rigidity over the course of the flight. 
Although the final fit to the flight data accounts for variations in the continuum spectra with changing rigidity, here we manually consider different rigidity regions.

\begin{figure}[htpb]
\centering
\includegraphics[width=1.0\columnwidth,trim=0.0in 0.0in 0.5in 0.3in, clip=true]{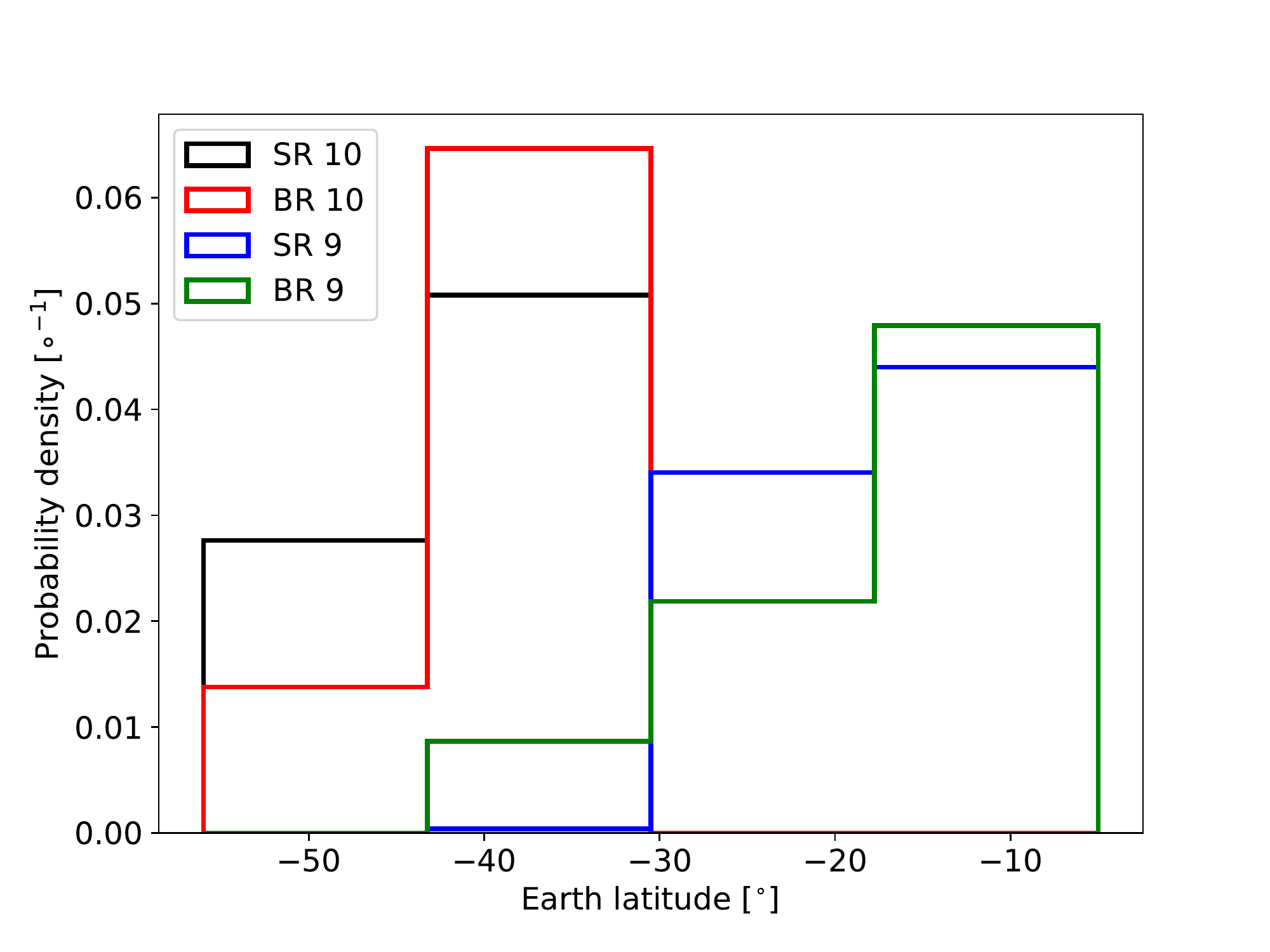}
\caption{COSI 2016 flight data in the signal and background regions from the 10- and 9-detector portions of the flight, binned by Earth latitude. The area under each distribution is normalized to 1.}
\label{fig:earth_latitude}
\end{figure}

Rigidity $R$ and latitude from Earth's magnetic equator $\lambda$ are related by $R = 14.5\rm{cos}^4(\lambda)/r^2$ \citep{smart2005review} for distance from Earth's dipole center r, regarded here as a constant.
As such, to account for rigidity we bin the signal region and background region flight data, each divided between the 10- and 9-detector portions of the flight, into four latitude bins (Figure\,\ref{fig:earth_latitude}).
We generate four energy spectra, each corresponding to one latitude bin, in the signal and background regions' 10- and 9-detector parts of the flight, i.e. 16 spectra total.   
We then re-weight the photon counts in the eight latitude spectra of the background region by the fraction of time COSI observed in the corresponding latitudes of the signal region (Figure\,\ref{fig:earth_latitude}).
After weighting, the four latitude spectra in each of the signal and background data sets are summed to form one energy spectrum, integrated over latitude, and combined over the 10- and 9-detector parts of the flight.   
Both spectra are normalized by the observation time in each region. 

\begin{figure}[htpb]
\centering
\includegraphics[width=1.0\columnwidth,trim=0.4in 0.1in 0.95in 0.7in, clip=true]{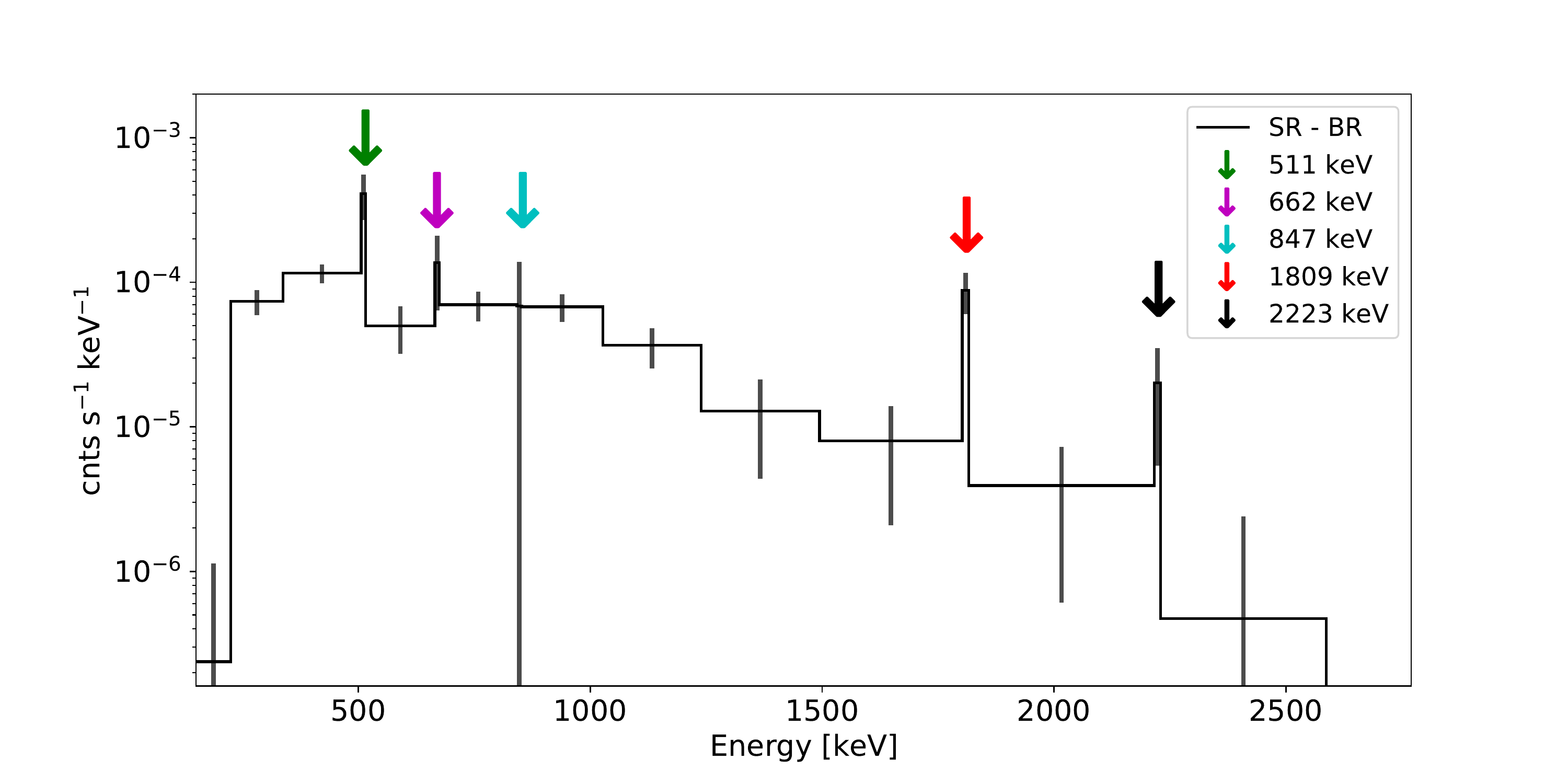}
\caption{Background-subtracted COSI 2016 spectrum of the signal and background regions after weighting the flight data by latitude, i.e. geomagnetic rigidity. Error bars are $\sqrt{\mathrm{counts}}$.}
\label{fig:rigidity_diff_flight_spectrum}
\end{figure}

The subtracted spectrum of the signal and weighted background region data is shown in Figure\,\ref{fig:rigidity_diff_flight_spectrum}.
After weighting by latitude (and thus rigidity), the 1809\,keV signature of \nuc{Al}{26} is clearly visible.
Some of the line features in the full flight spectrum (Figure\,\ref{fig:full_flight_spectrum}) disappear and the continuum is more suppressed. 
In particular, the $\sim$847\,keV line seen in Figure\,\ref{fig:full_flight_spectrum} is no longer visible.
We fit the spectrum to estimate the count rates of the remaining lines (Table\,\ref{table:rigidity_diff_flight_table}); those of instrumental origin are interpreted as systematic uncertainties in the analysis. 
The 511\,keV significance is smaller than that of 1809\,keV because the analysis is optimized to identify the 1809\,keV line.
Overall, the instrumental lines at 662\,keV, 847\,keV, and 2223\,keV are insignificant compared to 511\,keV and 1809\,keV.

\begin{table}
\centering
\begin{tabular}{ccc}
\hline
{Line energy} & {Integrated count rate} & {Significance} \\
{[keV]} & [10$^{-4}$ cnts s$^{-1}$] & {} \\
\hline
511  & 32 $\pm$ 11 & 2.9$\sigma$ \\
662  & 48 $\pm$ 42 & 1.1$\sigma$ \\
847  & 1.4 $\pm$ 5.8 & 0.2$\sigma$ \\
1809 & 8.3 $\pm$ 2.1 & 4.0$\sigma$ \\
2223 & 2.0 $\pm$ 1.2 & 1.7$\sigma$ \\
\hline
\end{tabular}%
\caption{Line rates and uncertainties after the rigidity-weighted subtraction.}
\label{table:rigidity_diff_flight_table}
\end{table}

After weighting by rigidity, we measure the \nuc{Al}{26} signal with 3.9$\sigma$ and find an Inner Galaxy flux of (10.7 $\pm$ 3.0) $\times$ 10$^{-4}$\,\flux.
This is consistent with previous iterations of the analysis.

\subsubsection{Broader energy range}
To demonstrate that our method can accommodate the continuum background independent of line emission, we expand the energy range of the analysis to 1650--1950\,keV.
We simulate the sky model over this new energy range as described in Sect.\,\ref{sec:sky_model} and empirically fit the new background region spectrum with a power law and five Gaussian-shaped lines.
We simultaneously fit these new models to the signal region data between 1650--1950\,keV and measure the \nuc{Al}{26} signal with 4.1$\sigma$ significance and an Inner Galaxy flux of (8.9 $\pm$ 2.4) $\times$ 10$^{-4}$\,\flux.
The slightly higher significance may indicate that by expanding the energy range, we are able to more strongly constrain the continuum in favor of the line signal. 
The consistency with the results in Sects.\,\ref{sec:signal_region} and \ref{sec:line_parameters} is affirmation of our method.

\subsubsection{Systematic uncertainties in flight data analysis}
\label{sec:flight_data_systematics}

The results from the previous tests of method validation are summarized in Table\,\ref{table:flight_method_validation}.
All Inner Galaxy flux values are consistent with each other within 1$\sigma$ uncertainties. 
They range from (3.8--13.7) $\times$ 10$^{-4}$\,\flux , placing a $\sim$57\% systematic uncertainty on the (8.6 $\pm$ 2.5) $\times$ 10$^{-4}$\,\flux measurement reported in Sect.\,\ref{sec:line_parameters}.  
Instrumental lines of less than 2$\sigma$ significance (Table\,\ref{table:rigidity_diff_flight_table}) indicate that the instrumental background is noticeably, if imperfectly, suppressed compared to lines of interest. 
Additional considerations of systematic uncertainties are derived from simulations in Sect.\,\ref{sec:sim_analysis} and a cumulative discussion of these uncertainties is presented in Sect.\,\ref{sec:systematic_uncertainties}.

\begin{table}
\centering
\begin{tabular}{ccc}
\hline
{Test}  & {Measurement} & {Inner Galaxy flux} \\
{} & {significance} & [10$^{-4}$\,\flux] \\
\hline
COMPTEL 1.8\,MeV & 3.6$\sigma$ & 6.6 $\pm$ 1.9\\
SPI 1.8\,MeV & 3.7$\sigma$ & 7.3 $\pm$ 2.1\\
M.A. Signal & 2.4--3.6$\sigma$ & 3.9--11.1\\
M.A. Background & 3.7$\sigma$ & 8.3 $\pm$ 2.5\\
Only 10-det. data & 2.3$\sigma$ & 6.8 $\pm$ 3.0\\
Only 9-det. data & 2.0$\sigma$ & 8.1 $\pm$ 4.1\\
Rigidity & 3.9$\sigma$ & 10.7 $\pm$ 3.0\\
1650--1950\,keV & 4.1$\sigma$ & 8.9 $\pm$ 2.4\\
\hline
\end{tabular}%
\caption{Summary of flight data results from various tests of method validation (Sect.\,\ref{sec:flight_method_validation}). ``M.A. Signal:" Minimum 27--33\,km altitudes in the signal region. ``M.A. Background:" Minimum 33\,km altitude in the background region.}
\label{table:flight_method_validation}
\end{table}

%\section{\texorpdfstring{\nuc{Al}{26} analysis with simulations}{26Al analysis with simulations}}
\section{Validating the method with simulations}
\label{sec:sim_analysis}
To further validate our method and results, the analysis outlined above is repeated on purely simulated data sets using four different template maps to model the \nuc{Al}{26} signal:
DIRBE $240\,\mrm{\mu m}$, SPI 1.8\,MeV, COMPTEL 1.8\,MeV, and ROSAT 0.25\,keV \citep{rosat}.
The latter is included as a map which traces high-latitude rather than Galactic Plane emission, and serves as a test of the sensitivity of our method.
We develop a simulated background model (Sect.\,\ref{sec:complete_flight_simulation}) and simulate COSI 2016 flights at different flux levels above this same background. We cross-check our results with statistical expectations (Sect.\,\ref{sec:various_DIRBE_sims}).
Finally, in Sect.\,\ref{sec:sim_analysis_onlyBG}, we perform an analysis on a data set comprised entirely of background as a measure of systematic uncertainty and validation of the real signal significance.

\begin{figure}[htpb]
\centering
\includegraphics[width=1.0\columnwidth,trim=0.0in 0.0in 0.5in 0.3in, clip=true]{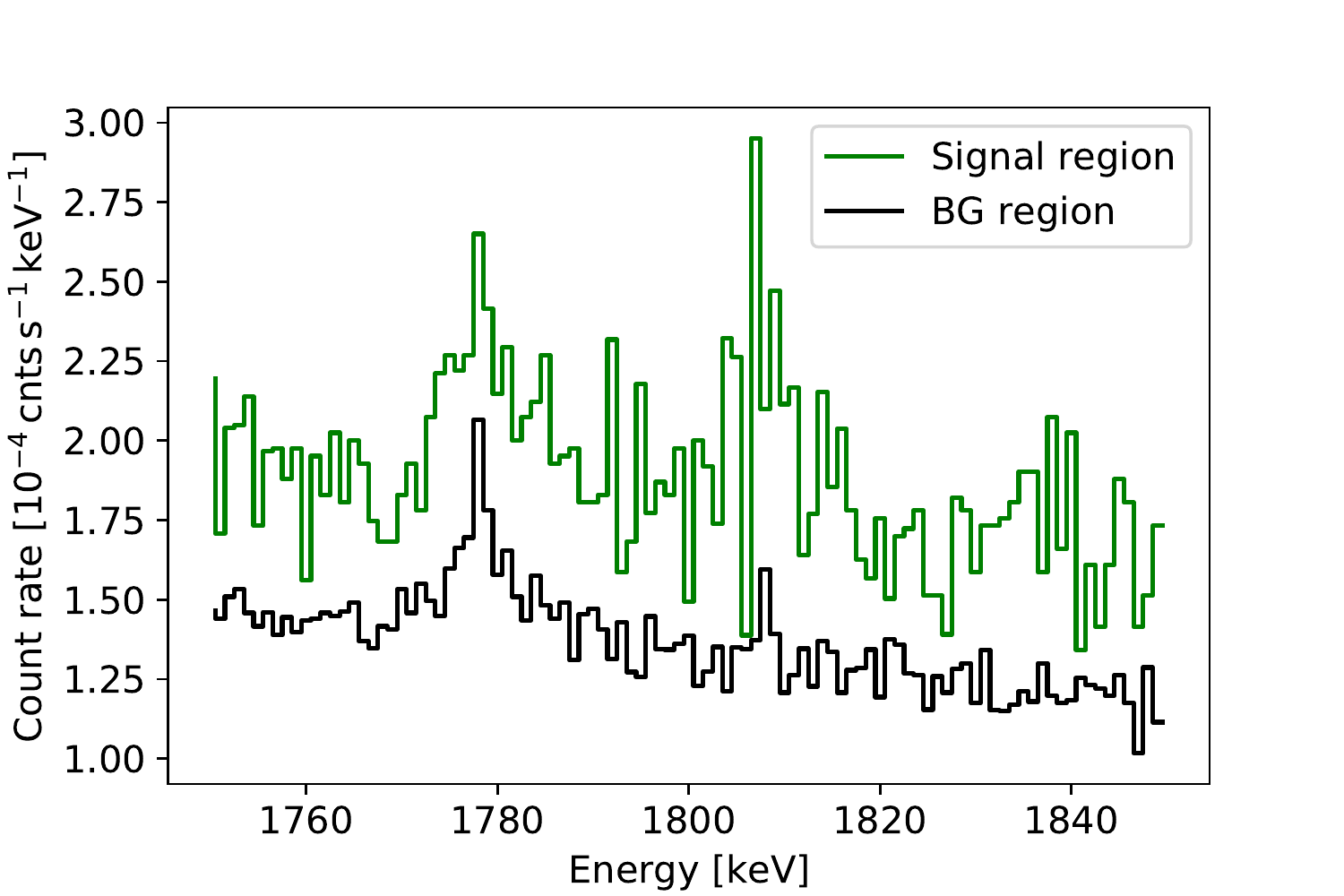}
\caption{Combined simulations of one 2016 flight over the DIRBE $240\,\mrm{\mu m}$ template image, instrumental activation background, and photonic background in signal and background regions, similar to Figure\,\ref{fig:flight_spectra}.}
\label{fig:sim_on_spec_off_spec_per_time}
\end{figure}

\subsection{Simulated data sets}\label{sec:sim_data_sets}
The simulations of the template maps are conducted assuming a constant transmission probability of $\sim 69.5\%$ at zenith (Figure\,\ref{fig:zenith_resp}), corresponding to a flight altitude of 33\,km.
The 10- and 9-detector portions of the flight are simulated separately with appropriate mass models. 
These simulations are run using MEGAlib's \citep{megalib} simulation tool called \texttt{cosima}, which is based on Geant4 \citep{1610988,ALLISON2016186,AGOSTINELLI2003250}. 
The template map simulations are combined with a \texttt{cosima} simulation of instrumental activation over 46 days of cosmic-ray and atmospheric particle irradiation \citep{activation} and a photonic background model to account for the Earth albedo \citep{ling_model}.
We scale the level of our background simulations to the best possible match with our flight observations.
We maintain the spectral shape of the simulated background.
The activation and photonic simulations together comprise the total simulated background and are discussed in more detail in Appendices \ref{sec:appx_activation_sims} and \ref{sec:appx_ling_sims}.
We apply the same pointing cuts and event selections from the flight data (Tables\,\ref{table:pointing_cuts} and \ref{table:event_selections}) to the combined signal and background simulated data sets.
This yields representative realizations of the COSI 2016 flight in the signal and background regions with a response to different \nuc{Al}{26} tracers.

\subsection{Complete flight simulation}\label{sec:complete_flight_simulation}
The simulated spectra in the signal and background regions of the DIRBE $240\,\mrm{\mu m}$ template image are shown in Figure\,\ref{fig:sim_on_spec_off_spec_per_time}.
These simulated spectra are similar to the flight spectra in Figure\,\ref{fig:flight_spectra}, suggesting a sufficiently accurate description of the data.
The background model is informed by applying minimal event selections to the combined activation and photonic simulations and fitting them with a power law and three Gaussian-shaped lines, Eq.\,(\ref{eq:bg_model}). This procedure is analogous to that with real flight data in Sect.\,\ref{sec:bg_model}. The simulated spectrum and fit parameters are shown in Figure\,\ref{fig:sim_fitted_plaw3Gauss_9and10dets} and Table\,\ref{table:sim_bg_priors}.

The largest differences in the simulated background spectrum are the count rates of the 1764 and 1779\,keV lines:
While in the flight data, the 1764\,keV is prominently seen, the activation simulation appears to show no 1764\,keV line at all.
This may be expected, however, because it is likely a line originating from the natural \nuc{U}{238} decay series, i.e. it is not due to local activation by cosmic-rays (Appendix\,\ref{sec:appx_activation_sims}).
The simulated 1779\,keV line appears as a blend of two lines at 1778 and 1784\,keV.
The slope of the background continuum is less steep around 1.8\,MeV with $\gamma_{\rm sim} \sim -3.7$ compared to $\gamma_{\rm flight} \sim -5.8$.
These differences motivate our empirical approach in the analysis of real flight data and underscore the difficulty of modeling the MeV background in a balloon environment.
As with the real flight data, the fitted spectral parameters of the simulated background and its uncertainties are fed as normal priors to a simultaneous fit of the sky and background models to the simulated signal region data.

\begin{figure}[htpb]
\centering
\includegraphics[width=1.0\columnwidth,trim=0.0in 0.0in 0.0in 0.0in, clip=true]{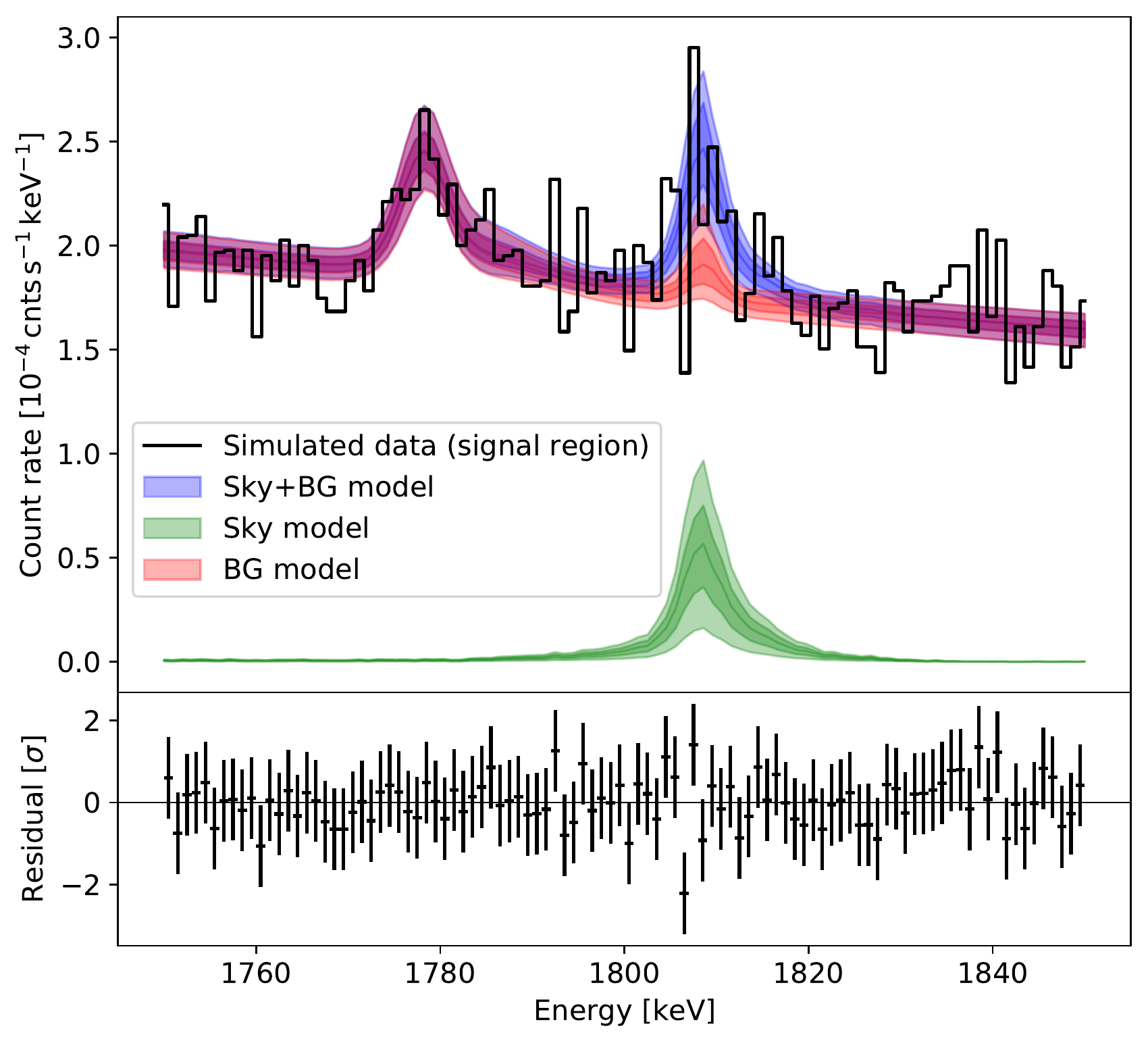}
\caption{Summed (Sky+BG) and individual sky and background models plotted over the signal region spectrum in the complete flight simulation, similar to Figure\,\ref{fig:flight_data_total_models}. Energy shift and broadening parameters are not considered in this figure, as we do not expect these astrophysical effects in simulated data.}
\label{fig:sim_total_models}
\end{figure}

The best-fit sky amplitude $\alpha = 0.7 \pm 0.3$ and the background amplitude $\beta = 28.7 \pm 0.6$.
The signal-to-noise ratio is $0.7/0.3 \sim 2.3$.
We note that this is less than the measured signal-to-noise ratio of $\sim 3.7$ in the real flight data.
We calculate a $2.8\sigma$ significance over the background compared to $3.7\sigma$ significance in the flight data.
The simulated signal rate between 1750 and 1850\,keV is $4.5 \times 10^{-4}\,\mrm{cnts\,s^{-1}}$, corresponding to $\sim 70$ \nuc{Al}{26} photons.
The simulated background region rate between 1803 and 1817\,keV is $2.9 \times 10^{-4}\,\mrm{cnts\,s^{-1}}$, corresponding to $\sim 392$ background photons.
Comparing to the real flight data, the simulated and flight background counts are comparable and the simulated sky photons are lower by a factor of $\sim 1.5$.
This difference suggests a systematic uncertainty in the absolute calibration of COSI's effective area (see Sect.\,\ref{sec:systematic_uncertainties}). 

\begin{figure}[htpb]
\includegraphics[width=1.0\columnwidth,trim=0.0in 0.0in 0.6in 0.7in, clip=true]{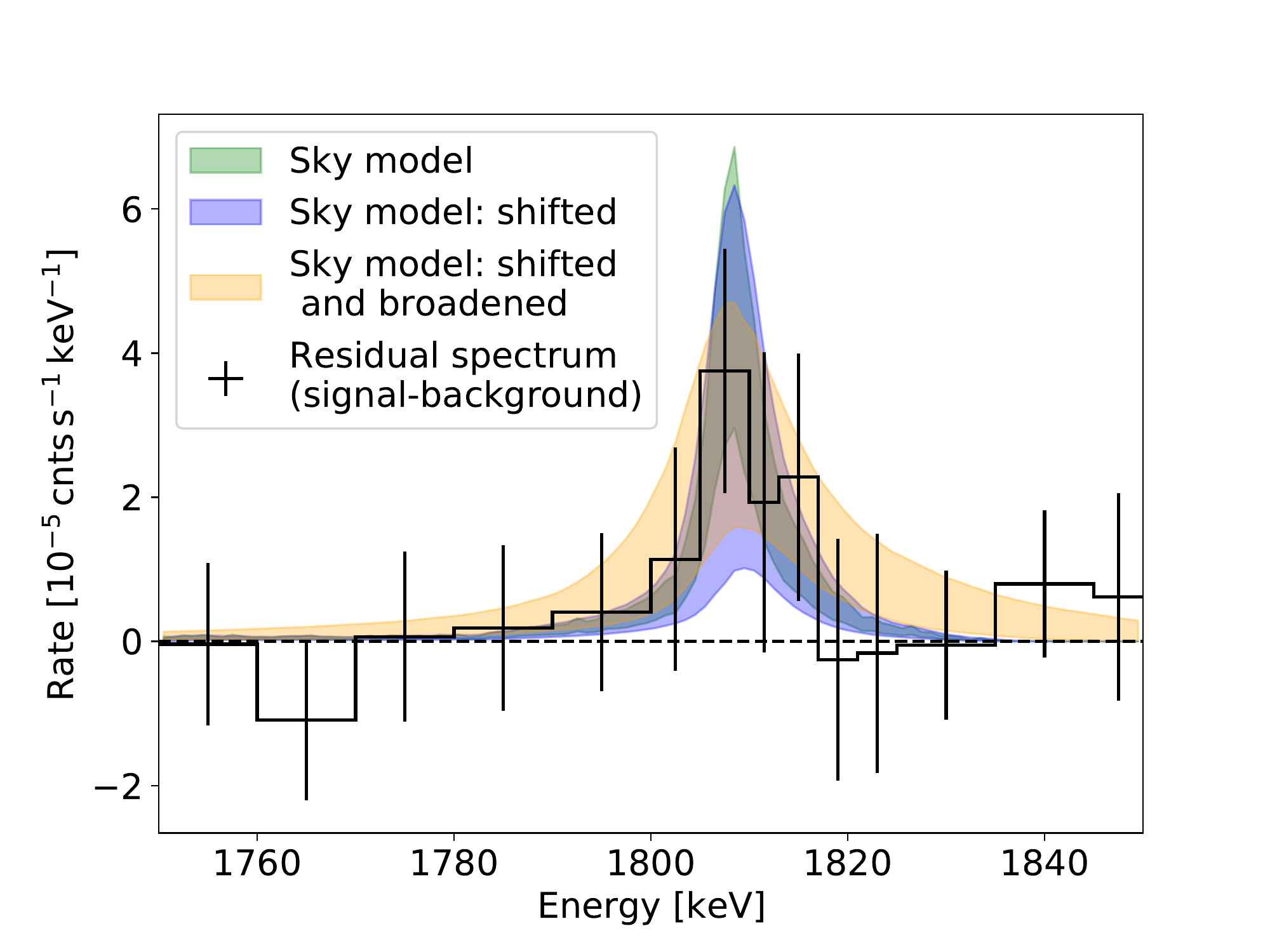}
\caption{Background-subtracted spectrum and 1$\sigma$ sky model contours from the complete flight simulation, similar to Figure\,\ref{fig:flight_data_bg_sub_spec}. Fitting for an energy shift gives $\Delta E$ = $-$0.2 $\pm$ 2.2\,keV. Fitting for line shift and broadening gives $\Delta E$ = 1.5 $\pm$ 1.7\,keV and an intrinsic sky broadening $<$ 13.7\,keV (2$\sigma$ upper limit).}
\label{fig:sim_bg_sub_spec}
\end{figure}

We plot the fitted total, sky, and background models for this simulation in Figure\,\ref{fig:sim_total_models}.
The background-subtracted spectrum is shown in Figure\,\ref{fig:sim_bg_sub_spec}.
The estimated \nuc{Al}{26} Inner Galaxy flux from this simulated data set is $(2.4 \pm 1.0) \times 10^{-4}$\,\flux.
Within uncertainties, this flux appears to be about $1.8$ times smaller than that of the flight data.
We also see a similar factor in the significance estimate, again suggesting a systematic offset. 

As with the flight data (Sect. \ref{sec:line_parameters}), we fit for an energy shift in the line. 
We expect $\Delta E$ to be consistent with zero because the simulated data do not include the intrinsic broadening of the sky seen in real flight data. 
Indeed we find a shift of $\Delta E = -0.2 \pm 2.2$\,keV and the Inner Galaxy flux is unchanged.
Including free parameters for shifting and broadening gives a shift of $\Delta E = 1.5 \pm 1.7$\,keV and a 2$\sigma$ upper limit on the intrinsic sky broadening of 13.7\,keV. 
The 1$\sigma$ contours of these shifted and broadened sky models are shown in Figure\,\ref{fig:sim_bg_sub_spec}.

\begin{table*}[t]
\tabcolsep=0.22cm
\centering
\begin{tabular}{cccccc}
\hline
Template map & Significance & Measured IG flux & Map IG flux & Sky amplitude & BG amplitude \\
& [$\sigma$] & [10$^{-4}$\,\flux] & [10$^{-4}$\,\flux] & $\alpha$ & $\beta$ \\
\hline
DIRBE $240\,\mrm{\mu m}$ & 2.8 $\pm$ 0.5 & 2.5 $\pm$ 0.4 & 3.3 & 0.7 $\pm$ 0.1 & 28.7 $\pm$ 0.1 \\
SPI 1.8\,MeV & 2.8 $\pm$ 0.4 & 1.9 $\pm$ 0.3 & 2.7 & 0.8 $\pm$ 0.1 & 28.8 $\pm$ 0.1 \\
COMPTEL 1.8\,MeV & 3.2 $\pm$ 0.5 & 2.5 $\pm$ 0.4 & 3.3 & 0.9 $\pm$ 0.1 & 28.8 $\pm$ 0.1 \\
ROSAT 0.25\,keV & --- & 0.2 $\pm$ 0.1 & 0.3 & 0.3 $\pm$ 0.1 & 28.7 $\pm$ 0.1 \\
\hline
\end{tabular}
\caption{Mean significance above background, measured \nuc{Al}{26} Inner Galaxy (IG) flux, true simulated \nuc{Al}{26} IG flux, sky amplitude $\alpha$, and background amplitude $\beta$ over 50 independent complete flight simulations of each tested map (Sect.\,\ref{sec:tracer_maps}).} 
\label{table:50diffsims_oneflight}
\end{table*}

\subsection{Simulations with different template maps}\label{sec:tracer_maps}
We repeat the analysis of Sect.\,\ref{sec:complete_flight_simulation} using the SPI 1.8\,MeV, COMPTEL 1.8\,MeV, and ROSAT 0.25\,keV images as template maps. 
Comparing the results across multiple template maps is both a check of the flight data measurement and a check of the consistency of our analysis pipeline.  

Table\,\ref{table:50diffsims_oneflight} shows the signal significance, measured \nuc{Al}{26} Inner Galaxy flux, true \nuc{Al}{26}  Inner Galaxy flux in the template map, and the best-fit $\alpha$ and $\beta$ averaged over 50 independent realizations of flight simulations per template map.
We find that the DIRBE, SPI, and COMPTEL template maps return Inner Galaxy fluxes consistent within two standard deviations of the true expected values.
The ROSAT map, which is not a tracer of \nuc{Al}{26} given its dearth of emission in the Inner Galaxy, yields a flux measurement nearly consistent with zero as expected.
This is affirmation of the null hypothesis:
the likelihood that COSI's signal region emission is traced by the ROSAT map is accounted for entirely by the background model ($\alpha \approx 0$, $\beta > 0$).

The analysis pipeline underestimates the \nuc{Al}{26} flux in the Inner Galaxy of each template map by about a factor of $1.5$.
This is probably due to the fact that the high latitude emission in the template maps is significantly different from zero.
The background model then absorbs some portion ($10$--$30$\%) of the total flux outside the Inner Galaxy.
In addition to the absolute effective area calibration, this value can be considered a systematic uncertainty in our method's definition of all emission outside of the Inner Galaxy as background (see Sect.\ref{sec:systematic_uncertainties} for further discussion). 
A better description of the \nuc{Al}{26} sky is necessary to constrain high-latitude emission and the resulting uncertainty.

\subsection{Increasing the signal}\label{sec:various_DIRBE_sims}
To assess the validity of our simulation, we conduct additional iterations of the analysis outlined above by simulating different flux levels above the simulated background.
To obtain an objective measure that our method works, we increase the flux in our simulations while keeping the background level constant.
That is, we pick at random $n$ out of 50 sky simulations and perform the same analysis as above to benchmark the simulation results against expectations.

\begin{figure}[htpb]
\includegraphics[width=1.0\columnwidth,trim=0.2in 0.0in 0.6in 0.5in, clip=true]{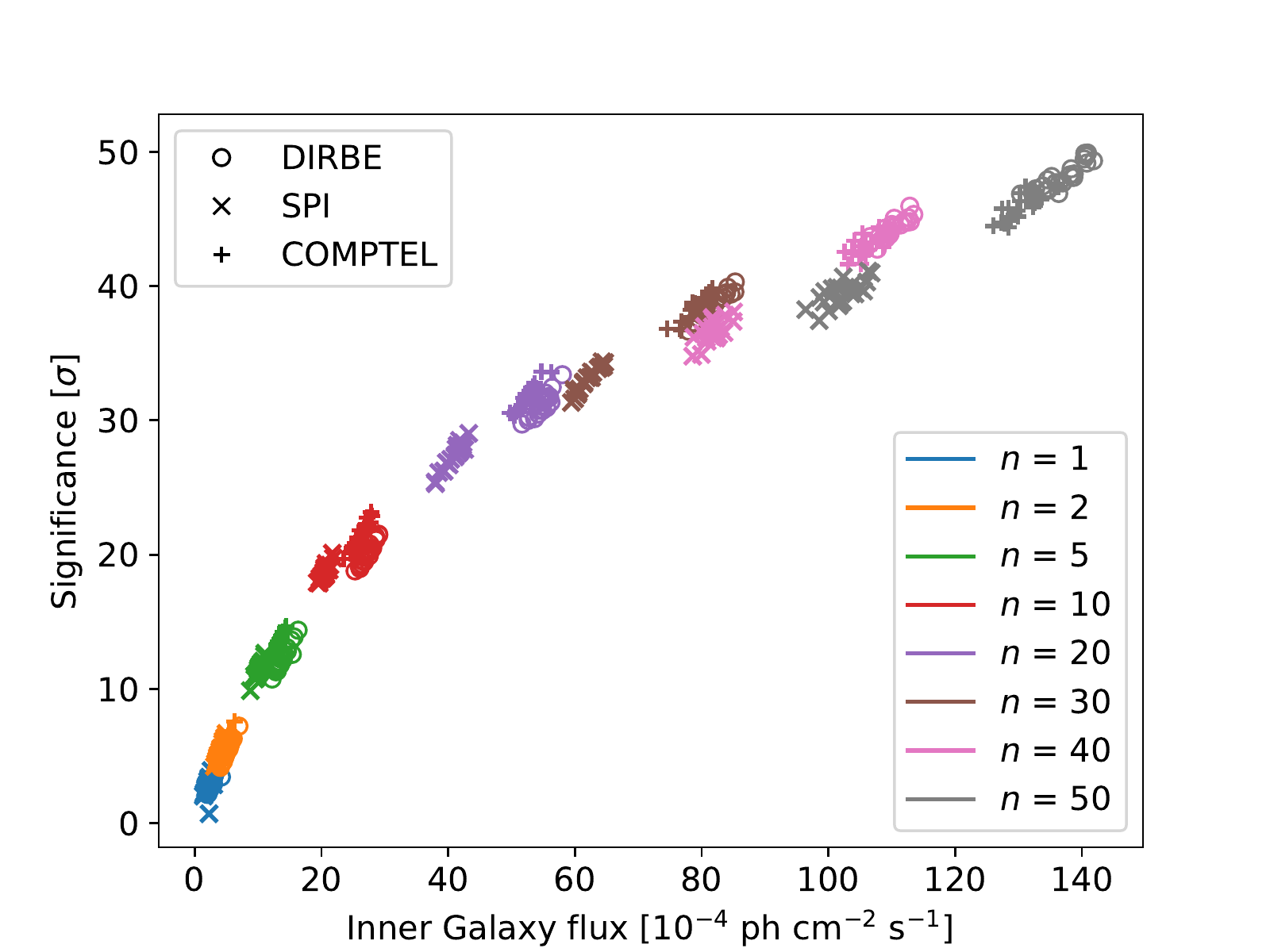}
\caption{Significance vs. estimated Inner Galaxy flux for simulated data sets containing $n$ DIRBE, SPI, and COMPTEL simulations of the flight combined with activation and photonic background simulations. The analysis was performed 25 times per $n$ simulations, indicating the scatter of different realizations.}
\label{fig:sim_analysis_randomcombosmaps_sig_vs_flux}
\end{figure}

For each case, we run 25 realizations by randomly selecting $n$ out of 50 simulations.
The background in each case is the simulated instrumental activation and photonic background described in Sect.\,\ref{sec:complete_flight_simulation}.
Figure\,\ref{fig:sim_analysis_randomcombosmaps_sig_vs_flux} shows the estimated significance against the estimated flux for the DIRBE, SPI, and COMPTEL maps.
We find the expected square-root-like behavior of increasing flux or, equivalently, exposure time.

As expected, using the ROSAT map as a template of \nuc{Al}{26} emission did not yield estimates of significant positive excess above the background.
This is further validation of the pipeline because the ROSAT map shows strong emission only at high latitudes.

\subsection{Background-only simulations}\label{sec:sim_analysis_onlyBG}
Finally, we repeat the analysis on simulated data sets devoid of any sky signal.
In this way, we obtain a distribution of test statistic (TS) values that follows a $\chi^2$-distribution with one degree of freedom, i.e. $\alpha = 0$ versus $\alpha \neq 0$ \citep[Wilks' theorem,][]{wilks1938large}.
We fit the background region spectrum from the flight data (Sect.\,\ref{sec:bg_model}) 1000 times.
In each iteration, we define the signal region spectrum as a Poisson sample of the flight data background model defined by the fit parameters describing the background spectrum.

\begin{figure}[htpb]
\includegraphics[width=1.0\columnwidth,trim=0.0in 0.0in 0.6in 0.4in, clip=true]{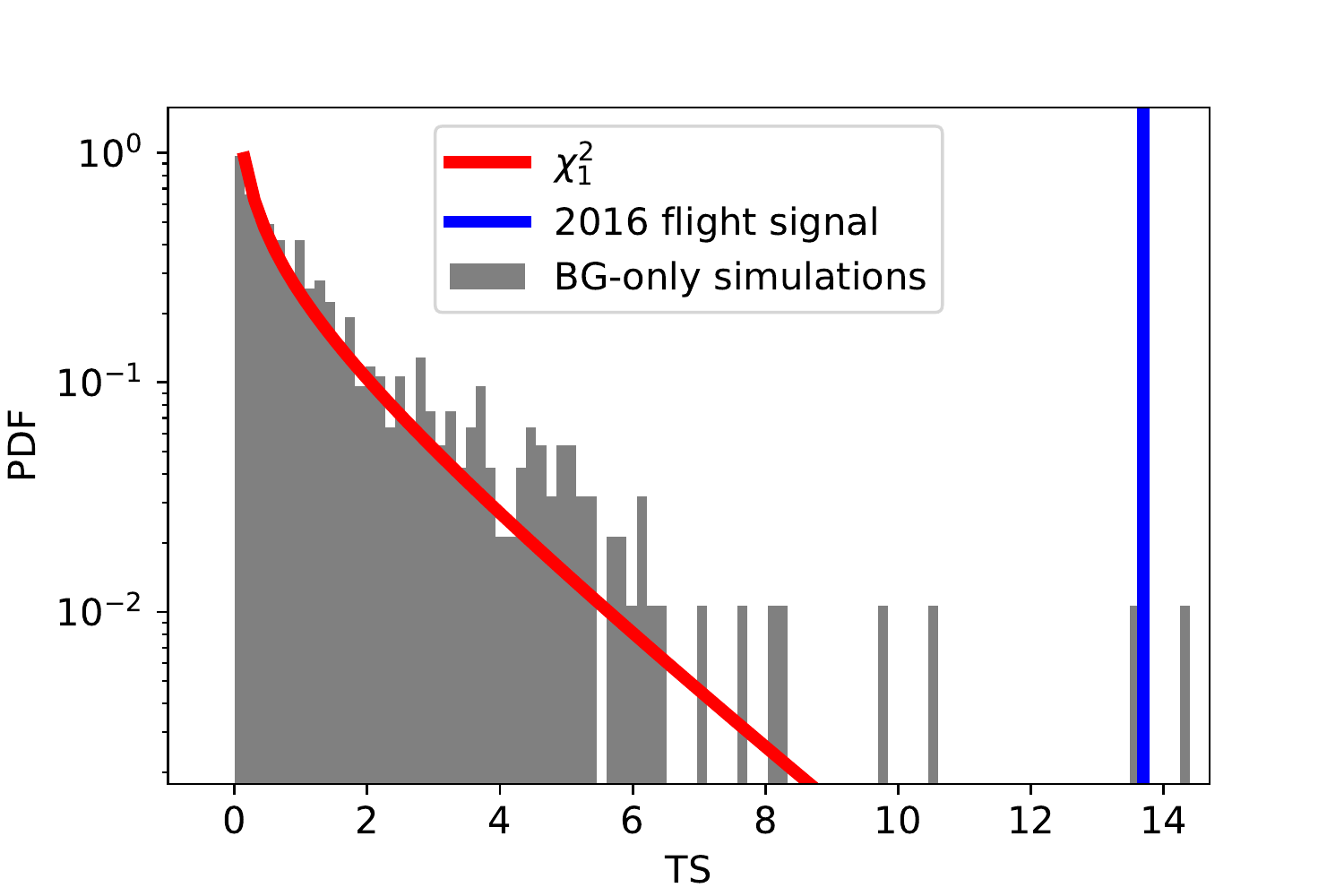}
\caption{Distribution of the test statistic (TS) from 1000 simulated data sets. The signal in each is defined as a Poisson realization of the fitted background model from flight data.}
\label{fig:bg_only_TS}
\end{figure}

Figure\,\ref{fig:bg_only_TS} demonstrates that the TS indeed follows a $\chi^2_1$-distribution.
The $3.7\sigma$ (equivalent to $p$ value $= 0.00022$) measurement from the real flight analysis clearly exceeds the significance returned by 1000 assumptions of the null hypothesis.
Thus, we verify that the TS calculated in our analysis method is a reliable proxy of the likelihood that the flight data $d$ are described by our model description $m$.

\section{Discussion}\label{sec:discussion}
\subsection{Comparison to previous measurements}\label{sec:comparison_prev_measurements}
Depending on the template map used, we find an \nuc{Al}{26} flux in the Inner Galaxy between $4.7 \times 10^{-4}$\,\flux and $11.1 \times 10^{-4}$\,\flux.
Our measured flux is consistent with previous measurements from SPI and COMPTEL of $2.8$--$3.3 \times 10^{-4}$\,\flux within $2\sigma$ uncertainties.
We find a line centroid of $1811.2 \pm 1.8$\,keV using the DIRBE, SPI, and COMPTEL template maps. This is consistent with previous measurements and in particular with the laboratory energy of 1808.7\,keV within $2\sigma$ uncertainties.
While SPI measured a Doppler shift of $1809.02 \pm 0.04$\,keV in the Inner Galaxy \citep{siegert2017positron}, the systematic uncertainties in these measurements due to calibration, detector degradation, and line shape are about one order of magnitude larger than the statistical uncertainties.
We repeat the COSI flight analysis in Sect.\,\ref{sec:line_parameters} with the line shift fixed to 0.3\,keV (to the SPI centroid of $1809.02 \pm 0.04$\,keV). 
This gives an Inner Galaxy flux of $(9.9 \pm 2.8) \times 10^{-4}$\,\flux, which is fully consistent with the results when the line shift is left as a free parameter. 
Overall, the absolute line shift in the Inner Galaxy is difficult to model because individual stellar groups, the large-scale Galactic rotation, and preferential streaming directed along Galactic rotation \citep{Kretschmer_2013} all contribute to the total line shift.

Our line width places a 2$\sigma$ upper limit on the turbulent motion of \nuc{Al}{26} ejecta in the Inner Galaxy of $\lesssim 2800$\,\vel.
Accounting for the large scale motion as measured in \citet{Kretschmer_2013}, the intrinsic velocity broadening is limited to $\lesssim 2400$\,\vel.
This is about one order of magnitude greater than the expected turbulent motion of the hot gas in the ISM, where a line width of 1\,keV corresponds to a velocity of 122\,\vel \citep{diehl2006radioactive,Wang2009_26Al}.
In 1996, the balloon-borne Gamma Ray Imaging Spectrometer (GRIS) also reported a wide intrinsic sky broadening of 5.4$^{+1.4}_{-1.3}$\,keV and a velocity $>$ 450\,\vel, which exceeds expectations from motion of hot gas in the ISM \citep{naya1996detection}.
The difficulty of measuring the broadening precisely is clear, despite the excellent energy resolution of germanium detectors. Adding an instrumental resolution of $\sim$ 3\,keV at 1809\,keV in quadrature with an intrinsic sky broadening of 1\,keV, for example, gives a measured line width of $\sim$ 3.2\,keV. The measured width in this toy example is only $\sim$ 7\% larger than the instrumental resolution, even though the intrinsic sky broadening is 33\% as wide as the instrumental resolution.

A measurement of the Galaxy-wide Doppler broadening of the 1.8\,MeV emission also remains an open issue because measuring the broadening, rather than the shift, requires considerably longer integration times.
Detectors degrade over these long integration times, changing the instrumental line response and complicating the analysis.
However, as a satellite mission, COSI-SMEX's enhanced line sensitivity of $1.7 \times 10^{-6}$\,\flux at 1809\,keV \citep[3$\sigma$ over 24-month survey,][]{tomsick2019compton} compared to INTEGRAL/SPI may expedite a Doppler broadening measurement of the \nuc{Al}{26} line.
Additionally, the satellite's improved angular resolution of 1.5$^{\circ}$ \citep{tomsick2019compton} has potential to advance explorations of \nuc{Al}{26} dynamics \citep{krause201526al,fujimoto2020distribution} and those of recently created elements \citep{forbes2021solar}.   

\subsection{Systematic uncertainties}\label{sec:systematic_uncertainties}
The \nuc{Al}{26} flux value measured in the COSI 2016 flight is approximately two times greater than expected.
This enhancement is similar to that seen in analyses of the 511\,keV positron annihilation line during the COSI flight \citep[cf.][]{siegert2020imaging,kierans2020detection}.
Applying this systematic factor to the \nuc{Al}{26} measurement gives an Inner Galaxy flux of $(4.3 \pm 1.3) \times 10^{-4}$\,\flux, consistent with previous measurements from SPI and COMPTEL.
Thus, we see a systematic uncertainty on the overall flux normalization $\sim$ 50\%, probably owing to the absolute calibration of the effective area, independent of energy.
This uncertainty may also be attributed to possible imperfections in the atmospheric model assumed by MEGAlib when simulating COSI's spectral sky model at a minimum altitude of 33\,km. 
Repeating the flight data analysis under a variety of conditions (Sect.\,\ref{sec:flight_method_validation}) also indicates a systematic uncertainty on the flux of $\sim$ 57\%.

Additional systematics arise from the analysis method itself.
Our approach relies on the assumption that at high latitudes ($|b| \gtrsim 45^{\circ}$) and longitudes ($|\ell| \gtrsim 105^{\circ}$), the sky is devoid of any \nuc{Al}{26} signal.
The template maps used for the signal, DIRBE $240\,\mrm{\mu m}$, SPI 1.8\,MeV, and COMPTEL 1.8\,MeV, all show a non-zero contribution in these background regions.
While we can estimate the flux contribution from regions like Orion, Perseus, Taurus, Carina, or Vela from previous studies to account for at most $15\%$ of the total \nuc{Al}{26} emission \citep[see, e.g.,][]{Bouchet_2015,siegert2017positron,moritz_thesis}, the emission at high latitudes is essentially unknown.
The COMPTEL map shows nearly-homogeneous diffuse emission at these latitudes, which is likely residual emission from the reconstruction algorithm.
Likewise, the SPI 1.8\,MeV image shows one particularly bright spot at $(\ell,b) = (226^{\circ}, 76^{\circ})$, which is almost certainly an artifact in the image reconstruction because no \nuc{Al}{26} source is known at this position with a flux of $5$--$9 \times 10^{-5}$\,\flux \citep{Bouchet_2015}.
Finally, because the DIRBE $240\,\mrm{\mu m}$ map performs well in a fit to raw $\gamma$-ray data from SPI and COMPTEL, it only traces, rather than shows directly, the true distribution of \nuc{Al}{26}.
We estimate the systematic uncertainties in the template map as $10$--$30\%$, given the DIRBE $240\,\mrm{\mu m}$ simulated flux of $(2.5 \pm 0.4) \times 10^{-4}$\,\flux (Table\,\ref{table:50diffsims_oneflight}) compared to the true map flux of $3.3 \times 10^{-4}$\,\flux.

We perform an additional check of this systematic by modifying the DIRBE $240\,\mrm{\mu m}$ template image to contain zero flux outside of the 35$^{\circ}$-broadened Inner Galaxy ($|\ell| \leq 65^{\circ}$, $|b| \leq 45^{\circ}$) and repeating the flight data analysis.
This artificial map, which contains \nuc{Al}{26} only in the broadened signal region, yields an Inner Galaxy flux of $(9.3 \pm 2.7) \times 10^{-4}$\,\flux.
The enhanced flux confirms that defining unconstrained emission of \nuc{Al}{26} at higher latitudes as background introduces systematic uncertainty.  
We also note that its consistency with the flight measurement of $(8.6 \pm 2.5) \times 10^{-4}$\,\flux is validation of the claim that COSI is sensitive to photons $\sim$35$^{\circ}$ beyond the Inner Galaxy.

This test may clarify the factor of $\sim 1.5$ seen in Sect.\,\ref{sec:sim_analysis} and clearly illustrates the need to constrain this systematic with a more detailed description of \nuc{Al}{26} across the entire sky. 
With the more unique imaging response of compact Compton telescopes compared to that of coded-mask instruments like SPI (which are not optimized for observing shallow emission gradients or isotropic emission), and better spectral resolution compared to NaI scintillators (COMPTEL), imaging high latitude emission is an achievable goal for COSI-SMEX.
Constrained high latitude emission will provide valuable insight to the open problem of the true \nuc{Al}{26} morphology in the Milky Way \citep{pleintinger2019comparing}.

\section{Summary}\label{sec:summary}
We report a 3.7$\sigma$ measurement of Galactic \nuc{Al}{26} in the COSI 2016 balloon flight. 
The Inner Galaxy ($|\ell| \leq 30^{\circ}$, $|b| \leq 10^{\circ}$) flux is estimated as ($ 8.6 \pm 2.5_{\rm stat} \pm 4.9_{\rm sys}) \times 10^{-4}$\,\flux. 
Within 2$\sigma$ uncertainties, this value is consistent with previous measurements by SPI and COMPTEL. 
Systematic uncertainties seen in previous COSI analyses of the 511\,keV positron annihilation line and those intrinsic to the assumption of no \nuc{Al}{26} emission at high latitudes may account for the discrepancy. 
We find a total line shift of $2.5 \pm 1.8$\,keV, an intrinsic line broadening of 9.7\,keV (2$\sigma$ upper limit), and limit the turbulent velocity of \nuc{Al}{26} ejecta to $\sim 2800$\,\vel (2$\sigma$ upper limit).
Extensive simulations of the flight with several template maps affirm the consistency of the analysis pipeline with expectations.
Overall, the framework behaves as expected and returns a 3.7$\sigma$ measurement above background, consistent with previous measurements within $\sim$2$\sigma$ uncertainties.

The COSI 2016 balloon flight's measurement of \nuc{Al}{26} is key proof-of-concept for future studies of nucleosynthesis. 
Its high-purity germanium detectors have excellent energy resolution ideal for $\gamma$-ray spectroscopy. 
Single-photon reconstruction and the unique imaging response of Compton telescopes are valuable assets to imaging studies. 
Advancing this technology to a satellite platform (COSI-SMEX) will strengthen the \nuc{Al}{26} balloon measurement and probe unsolved questions about its origin, distribution, dynamics, and influence on the early Solar System.
Preserving the advantages of germanium Compton telescopes as demonstrated in the balloon iteration, moving to low-Earth orbit presents a much more favorable background environment than the dominant atmospheric background and atmospheric attenuation seen in balloon missions \citep{cumani2019background}. 
These preferred background conditions and an additional layer of four germanium detectors will increase the effective area, thereby enhancing the observational capabilities of the satellite platform.
Thus, the next generation of MeV satellite missions, particularly Compton telescopes like COSI-SMEX, has potential to bring the MeV regime of $\gamma$-ray astrophysics into a new era of improved sensitivity and scientific understanding.

% Authors are encouraged to use an online tool at
% \url{http://authortools.aas.org/FIGSETS/make-figset.html} to generate their
% own specific figure set mark up to incorporate into their \latex\ articles.

%% IMPORTANT! The old "\acknowledgment" command has be depreciated. It was
%% not robust enough to handle our new dual anonymous review requirements and
%% thus been replaced with the acknowledgment environment. If you try to 
%% compile with \acknowledgment you will get an error print to the screen
%% and in the compiled pdf.
\begin{acknowledgments}
COSI is supported through NASA APRA Grant 80NSSC19K1389. This work is also supported in part by the Centre National d'\'Etudes Spatiales (CNES). Thomas Siegert is supported by the German Research Foundation (DFG-Forschungsstipendium SI 2502/3-1).
\end{acknowledgments}

%% Similar to \facility{}, there is the optional \software command to allow 
%% authors a place to specify which programs were used during the creation of 
%% the manuscript. Authors should list each code and include either a
%% citation or url to the code inside ()s when available.

\software{Astropy \citep{astropy:2013, astropy:2018}, emcee \citep{emcee}, matplotlib \citep{Hunter:2007}, MEGAlib \citep{megalib}, numpy \citep{harris2020array}, scipy \citep{2020SciPy-NMeth}.}

%% Appendix material should be preceded with a single \appendix command.
%% There should be a \section command for each appendix. Mark appendix
%% subsections with the same markup you use in the main body of the paper.

%% Each Appendix (indicated with \section) will be lettered A, B, C, etc.
%% The equation counter will reset when it encounters the \appendix
%% command and will number appendix equations (A1), (A2), etc. The
%% Figure and Table counter will not reset.

\appendix
\renewcommand\thefigure{\thesection.\arabic{figure}}
\renewcommand\theHfigure{\thesection.\arabic{figure}}
\renewcommand\thetable{\thesection.\arabic{table}}
\renewcommand\theHtable{\thesection.\arabic{table}}

\section{Additional materials}\label{sec:appendix_figures_tables}
\setcounter{figure}{0}
\setcounter{table}{0}

\begin{figure}[ht]
\centering
\includegraphics[scale=0.5]{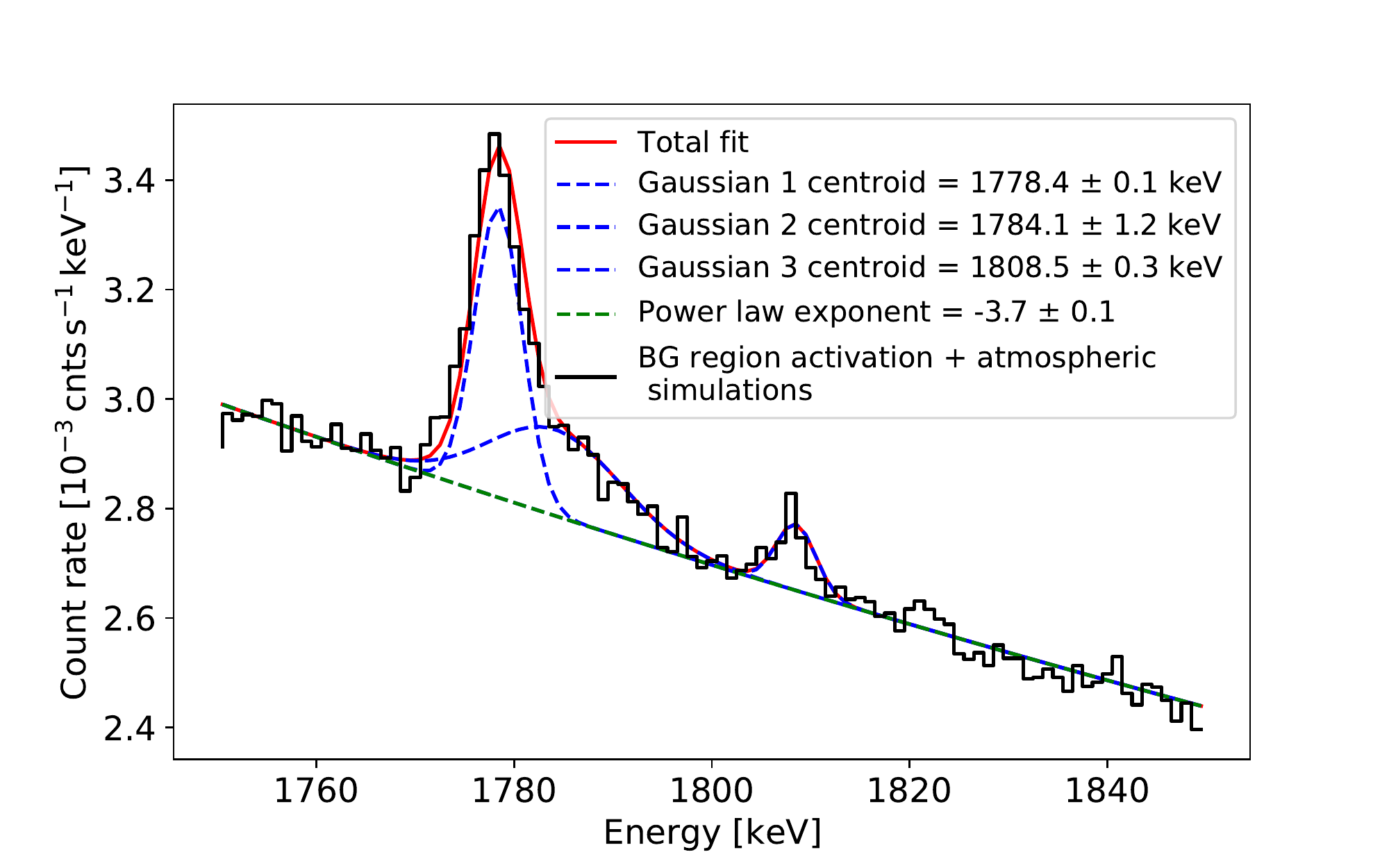}
\caption{Power law plus three Gaussian empirical fit to the instrumental activation and photonic background simulations with minimal event selections, similar to Figure\,\ref{fig:bg_model}. The parameters of the fit are listed in Table\,\ref{table:sim_bg_priors}.}
\label{fig:sim_fitted_plaw3Gauss_9and10dets}
\end{figure}

In Table\,\ref{table:flight_data_bg_priors} we list the parameters returned by an empirical fit of a power-law plus three Gaussian-shaped lines to the background region of the flight data (Figure\,\ref{fig:bg_model}).
In Table\,\ref{table:sim_bg_priors} we list the parameters of the simulated background spectrum, also fit with a power-law plus three Gaussian-shaped lines. The simulated spectrum is shown in Figure\,\ref{fig:sim_fitted_plaw3Gauss_9and10dets}.

\begin{table*}[htb]
\centerwidetable
\movetableright=-0.75in
\tabcolsep=0.32cm
\begin{tabular}{crrrrrrrrrrr}
\hline
& $C_0$  & $\gamma$ & $A_1$  & $E_1$  & $\sigma_1$ & $A_2$  & $E_2$ & $\sigma_2$ & $A_3$  & $E_3$  & $\sigma_3$ \\
\hline
Value       & 2.32 & -5.8     & 2.0 & 1763.8 & 3.8        & 5.2 & 1779.2 & 7.1        & 6.6 & 1808.0 & 6.6        \\
Uncertainty  & 0.03   & 0.3      & 0.7  & 0.8    & 1.0        & 0.8 & 0.6    & 1.2        & 0.6  & 1.0    & 0.5     \\
\hline
\end{tabular}
\caption{Fit parameters of a power law plus three Gaussian fit to the flight data in the background region with minimal event selections (Figure\,\ref{fig:bg_model}). Units: [$C_0]$ = $10^{-3}$ cnts s$^{-1}$ keV$^{-1}$, [$A_{l}$] = 10$^{-3}$ cnts s$^{-1}$, [$E_{l}$] = [$\sigma_{l}$] = keV.}
\label{table:flight_data_bg_priors}
\end{table*}

\begin{table*}[ht]
\tabcolsep=0.32cm
\centerwidetable
\movetableright=-0.75in
\begin{tabular}{crrrrrrrrrrr}
\hline
& $C_0$  & $\gamma$ & $A_1$  & $E_1$  & $\sigma_1$ & $A_2$  & $E_2$ & $\sigma_2$ & $A_3$  & $E_3$  & $\sigma_3$ \\
\hline
Value       & 2.69 & -3.7     & 3.2 & 1778.4 & 2.4        & 2.7 & 1784.1 & 6.7        & 0.6 & 1808.5 & 2.0        \\
Uncertainty  & 0.01   & 0.1      & 0.3  & 0.1    & 0.2        & 0.4 & 1.2    & 0.8        & 0.1  & 0.3    & 0.5     \\
\hline
\end{tabular}
\caption{Fit parameters of a power law plus three Gaussian fit to the simulated data in the background region with minimal event selections (Figure\,\ref{fig:sim_fitted_plaw3Gauss_9and10dets}). Units: [$C_0]$ = $10^{-3}$ cnts s$^{-1}$ keV$^{-1}$, [$A_{l}$] = 10$^{-3}$ cnts s$^{-1}$, [$E_{l}$] = [$\sigma_{l}$] = keV.} 
\label{table:sim_bg_priors}
\end{table*}

\section{\texorpdfstring{Optimization of Compton Scattering Angle $\mathrm{\phi}$}{Optimization of Compton Scattering Angle phi}}\label{appx:optimize_phi}
\setcounter{figure}{0}
\setcounter{table}{0}
To preferentially select \nuc{Al}{26} events over the abundant background events in both the signal and background regions, we employ a scanning procedure over the Compton scattering angle $\phi$ to identify an ideal range of allowed $\phi$-values in the signal and background spectra.
Identifying the maximum value also informs selection of the pointing cuts listed in Table\,\ref{table:pointing_cuts} which define the signal and background regions.
This $\phi_{\rm max}$ effectively broadens the region of the sky included in each pointing cut because photons recorded in each region may originate up to $\phi_{\rm max}$ outside of that region.
The signal region (the Inner Galaxy) is broadened by $\phi_{\rm max}$ to ($|\ell| \leq 30^{\circ} + \phi_{\rm max}$, $|b| \leq 10^{\circ} + \phi_{\rm max}$).
To avoid overlap between the signal and background regions, the latter is defined such that the extent of its $\phi_{\rm max}$-broadened border encloses everywhere outside of the broadened signal region.
Identifying the ideal minimum and maximum $\phi$ is discussed in this section.

Simulated \nuc{Al}{26} events define the signal data set for this optimization procedure.
The signal is generated via an all-sky simulation of COSI's response over the 2016 flight to \nuc{Al}{26} events traced by the DIRBE $240\,\mrm{\mu m}$ map.
The simulation is run for both the 10- and 9-detector flight configurations of the instrument.
The background data set for this procedure is simulated as atmospheric background photons on 2016 June 12 \citep{ling_model}.
On this day, COSI's altitude remained fairly stable near its nominal flight altitude of 33\,km and it had nine active detectors.
The high altitude on this day represents the best-case observing conditions for COSI in terms of mitigating the effects of Earth albedo and atmospheric absorption.
We use simulations for this optimization procedure rather than real data because the latter are subject to uncertainties and are always background-dominated, which prevents a clean comparison of Compton scattering cuts on the \nuc{Al}{26} versus background photons.

\begin{figure}[htpb]
\centering
  \includegraphics[width=0.45\linewidth]{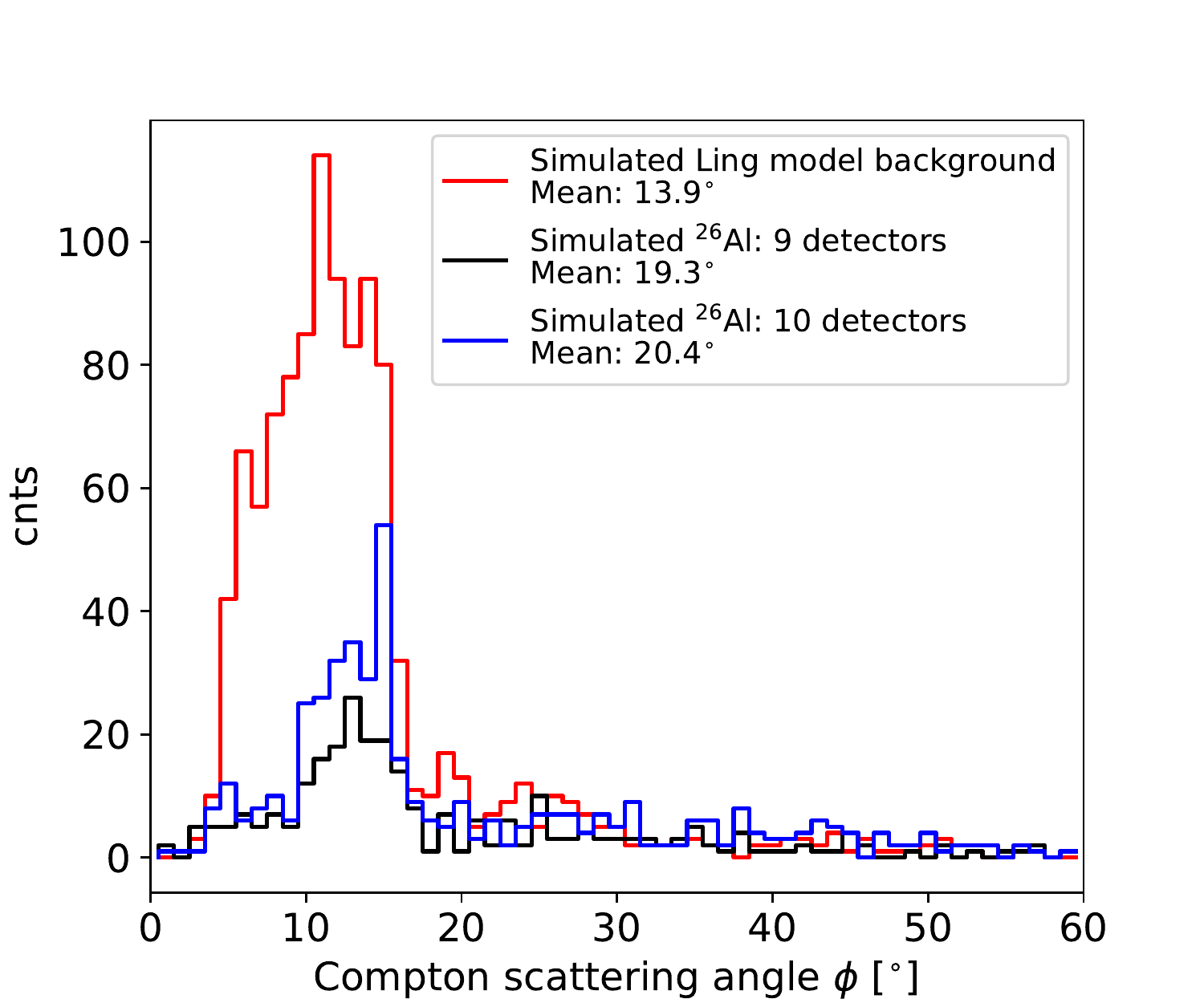}
    \includegraphics[width=0.45\linewidth]{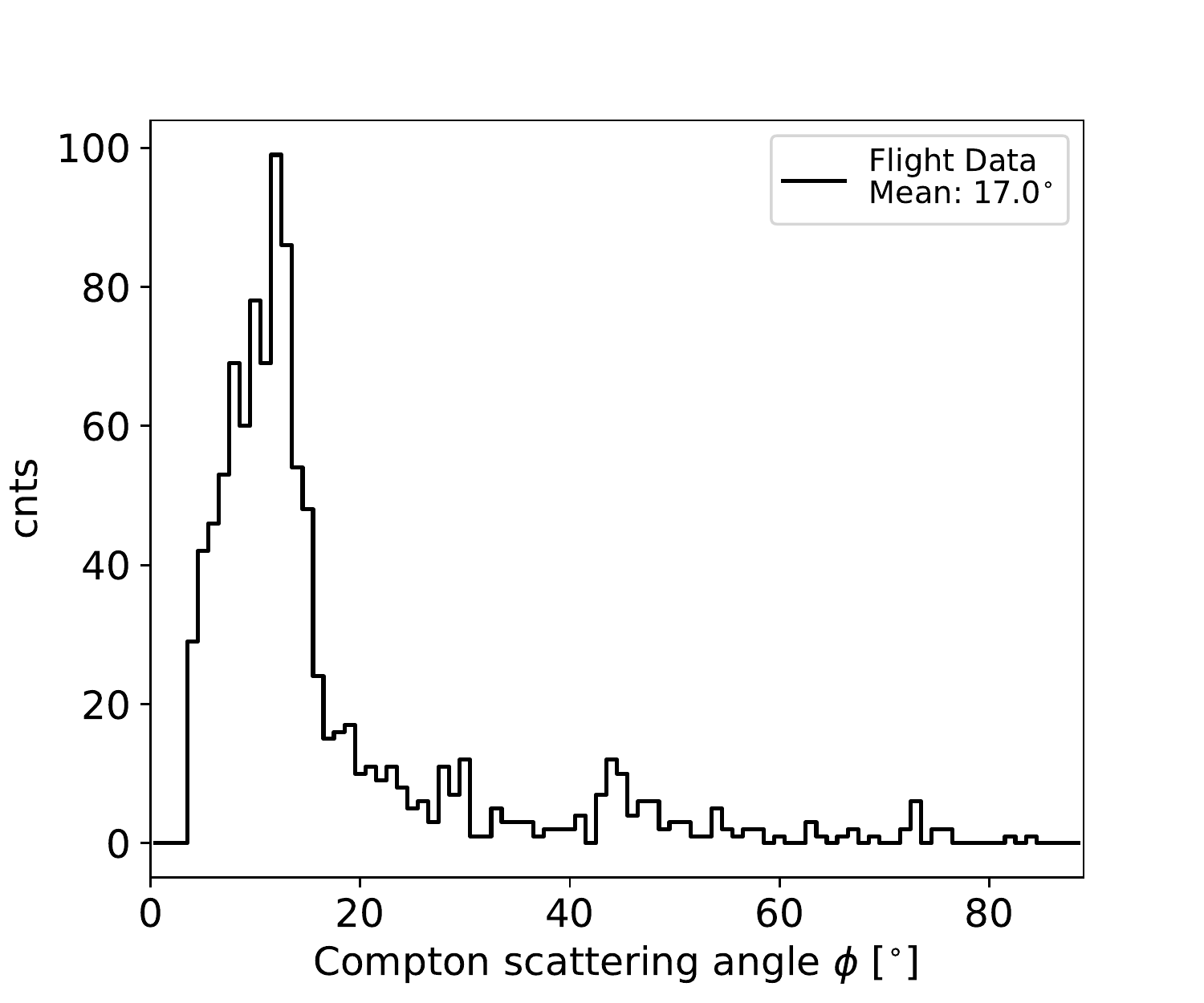}
\caption{Left: Distributions of Compton scattering angles from simulated \nuc{Al}{26} and background photons with incident energies 1803--1817\,keV. The \nuc{Al}{26} simulations are shown for both the 9- and 10-detector portions of the COSI 2016 flight. Right: Compton scattering angles from real flight data (1803--1817 keV; 10 detectors).}
\label{fig:phi_distribution_phi_step_down}
\end{figure}

The simulated photons are binned into one time bin spanning the flight time in their respective configurations (10 detectors: 2016 May 17 to 2016 June 5, 9 detectors: 2016 June 6 to 2016 July 2).
To focus on the energy band of interest for \nuc{Al}{26}, only events with incident energy between 1803 and 1817\,keV are analyzed.
We seek the range of allowed Compton scattering angles which rejects more background than celestial \nuc{Al}{26} events.
A histogram of $\phi$-values reveals that for both the \nuc{Al}{26} and background simulations, the large majority of events have $\phi$ less than 60$^{\circ}$ (Figure\,\ref{fig:phi_distribution_phi_step_down}).
Also visible in Figure\,\ref{fig:phi_distribution_phi_step_down}, which includes events with two or more interactions, is a sharp drop in events after $\sim$15$^{\circ}$. 
This drop is expected because the event reconstruction algorithm cannot deduce the order of interactions in many 2-site events. 
This means that the incident photon has two possible flight directions; these events are rejected from the analysis \citep{zoglauer2005first}. 
When combined with events of greater than two interactions, we see the effect in both the simulated and real flight data.

The background events appear more forward-scattered than the \nuc{Al}{26} events, despite the fact that the energy ranges in both are identically set to 1803--1817\,keV.
A plausible explanation for this discrepancy might be that a higher energy (background) photon, e.g. 5\,MeV, could deposit only 1.8\,MeV as it traverses the detector volume.
It then could escape detection without a final photo-absorption, carrying the remaining 3.2\,MeV and leaving behind a false 1.8\,MeV signature.
The hypothetical photon, with true energy greater than that recorded by COSI, would Compton scatter at smaller angles and skew the distribution to smaller values than those seen in true \nuc{Al}{26} events. 
We therefore examine the impact of changing the minimum and maximum allowable values of $\phi$ on the \nuc{Al}{26} and background events.

We recognize that a maximum Compton scatter angle cut of 60$^{\circ}$ yields the greatest overall number of \nuc{Al}{26} events simply because it permits the broadest possible range of allowed $\phi$ values.
However, allowing events from the signal region with such a high maximum $\phi$ effectively expands the signal region to occupy a significant portion of the total sky.
This leaves less space available for the remaining background region, resulting in fewer background events available to populate a robust background spectral template.
A well-determined background is important for minimizing uncertainties in later stages of the analysis.

\begin{figure}[htpb]
  \centering
  \includegraphics[width=0.8\linewidth]{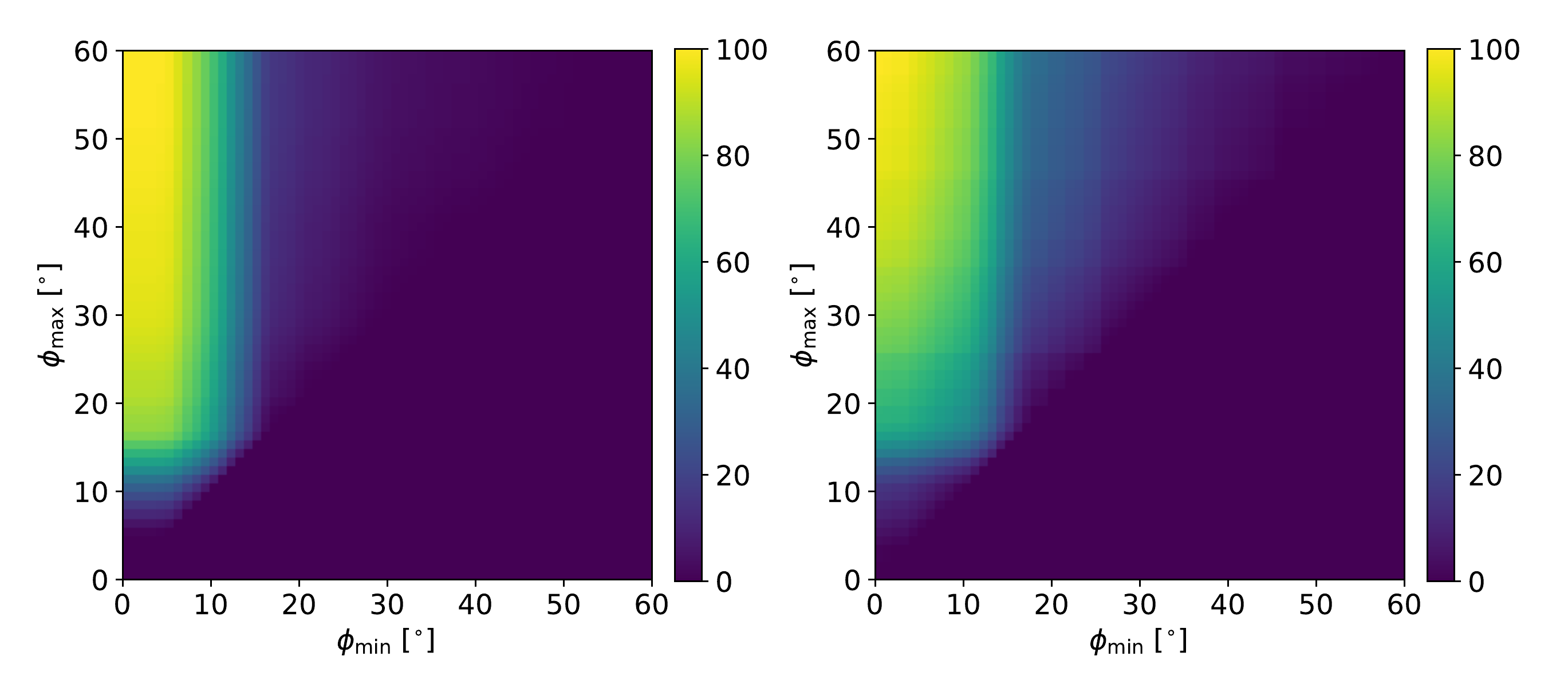}
  \caption{Compton scattering angle optimization procedure. The color scale indicates the percentage of events (left: Ling model background, right: \nuc{Al}{26} signal) that pass a cut allowing values of $\phi$ between the minimum (\textit{x}-axis) and maximum (\textit{y}-axis) limits, respectively. The maximum acceptance is seen with the broadest possible cut of 0$^{\circ}$ to 60$^{\circ}$. The increased forward scattering of background events (left) relative to the \nuc{Al}{26} events (right) is evident in the enhanced presence of background counts towards smaller scattering angles.}
  \label{fig:phi_simultaneous}
\end{figure}

Thus, for a more complete visualization of the impact of $\phi$ cuts on the \nuc{Al}{26} and background simulations, we optimize the lower and upper boundaries of $\phi$ simultaneously.
We probe every acceptable range of $\phi$ defined by minimum and maximum values each spanning 0--60$^{\circ}$.
Figure\,\ref{fig:phi_simultaneous} shows the percentage of events that pass a cut allowing values of $\phi$ between the minimum and maximum values.
The loosest cut of 0--60$^{\circ}$ accepts the most events, as expected, and the tendency of background events to undergo Compton scattering at smaller $\phi$ is evident in the enhanced presence of background counts towards smaller scattering angles.

\begin{figure}[htpb]
  \centering
  \includegraphics[width=0.6\columnwidth,trim=0.0in 0.0in 0.0in 0.0in,clip=true]{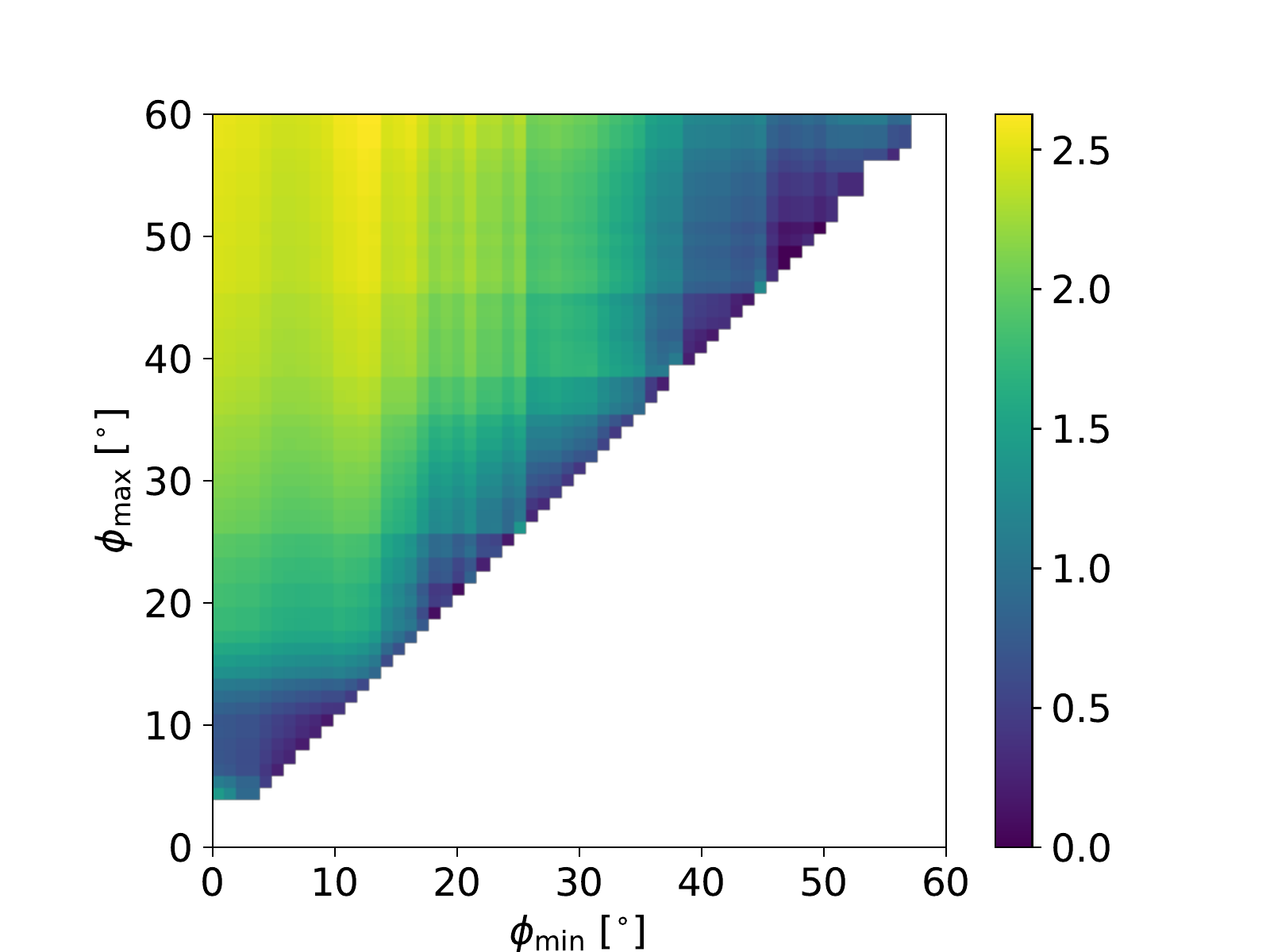}
  \caption{Estimated significance (signal / $\sqrt{\mathrm{background}}$) as a function of cuts in $\phi$ defined by the minimum and maximum values indicated on the axes. A $\phi_{\rm min}$ of $\sim 12^{\circ}$ and a $\phi_{\rm max} = 60^{\circ}$ yields the greatest $S/\sqrt{B} \sim 2.6$. In order to obtain suitable statistics for the background region without overlapping the signal region, we use a maximum Compton scattering angle of $35^{\circ}$.}
  \label{fig:phi_DIRBE_Ling_significance}
\end{figure}

As a gauge of signal-to-background significance, the raw numbers of events used to calculate the percentages in Figure\,\ref{fig:phi_simultaneous} are scaled to match the \nuc{Al}{26} and background simulations in flux.
The full-sky DIRBE map flux used in the simulations is $1.1 \times 10^{-3}$\,\flux and the most recent value from the literature is $1.7$--$1.8 \times 10^{-3}$\,\flux, so the \nuc{Al}{26} counts are scaled up by a factor of 1.6.

The results of this signal-to-background optimization procedure are shown in Figure\,\ref{fig:phi_DIRBE_Ling_significance}.
The significance is maximized at $\phi_{\rm min}$ = 12$^{\circ}$ and $\phi_{\rm max}$ = 60$^{\circ}$ with a value of $S/\sqrt{B} \sim 2.6$.
The maximum of 60$^{\circ}$ is always preferred because it yields the greatest number of \nuc{Al}{26} events, as explained above.
Setting the minimum to 10$^{\circ}$ rejects the domain of approximately 6$^{\circ}$ to 10$^{\circ}$, where the fraction of background dominates that of \nuc{Al}{26} events.

In choosing our final optimal $\phi$ cut we consider that we require sufficient statistics in the background region to obtain a robust background spectrum.
We finally choose to allow events with $\phi \in [10^{\circ}, 35^{\circ}]$.
The minimum of 10$^{\circ}$ accepts more \nuc{Al}{26} events than a minimum of 12$^{\circ}$ while still removing the background-heavy range of 6$^{\circ}$ to 10$^{\circ}$.
The maximum of 35$^{\circ}$, although quite restrictive, allows for a broader background region of the sky and, as shown in Figure \ref{fig:phi_DIRBE_Ling_significance}, yields an acceptable balance of \nuc{Al}{26} to background.
If a better standalone description of the instrumental background were available, the maximum Compton scattering angle could be relaxed to its optimum value, and the expected significance of the of \nuc{Al}{26} would be enhanced by $\sim 20\%$.

In Appendix\,\ref{appx:activation_ling_bg_sims}, we detail our efforts to build a more complete background model including atmospheric photons as well as activation from the instrument itself.
We show that although the levels and continuum shape of the background can be matched to some extent, the instrumental background lines in this energy range are difficult to model precisely.

\section{Instrumental activation and Atmospheric Background Simulations}\label{appx:activation_ling_bg_sims}
\setcounter{figure}{0}
\setcounter{table}{0}

\subsection{Activation Simulations}\label{sec:appx_activation_sims}
When cosmic rays and atmospheric particles strike the materials comprising and surrounding the COSI instrument, they have the potential to excite the nuclei in the materials to unstable states, which then de-excite and emit $\gamma$-rays.
These $\gamma$-rays can infiltrate the detectors and act as background to $\gamma$-rays from astrophysical sources of interest.
Hence, it is important to simulate the $\gamma$-rays from activation in order to understand the instrumental background in the data set. 

Activation simulations of various cosmic ray and atmospheric particles are performed in MEGAlib in three steps.
The dominant particle types are protons $p$, neutrons $n$, and $\alpha$-particles.
Emission from other particles, including muons, electrons, and positrons, was found to constitute a much smaller fraction of the background ($\sim$0.1\%) in previous activation simulations \citep{kierans2018}.
The first step (1) simulates the initial particles generated in the bombardment.
Prompt emission from these particles, meaning emission from excitations that decay on a timescale less than the detector timing resolution of $5\,\mrm{\mu s}$, and a list of all produced isotopes are stored.
This list of isotopes is the input to step (2) of the simulations, which calculates the activation of each isotope after a specified irradiation time.
The final step (3) of the activation simulations yields the delayed emission from the decays and de-excitations of extended irradiation encoded in step (2). 

Step (1) of each particle type was performed by \citet{kierans2018}.
For the purposes of this article, an irradiation time of 23 days is chosen for step (2) of the simulations to examine activation halfway through the COSI 2016 flight.
Step (3) is run for 46 days in order to approximate the full activation background over the COSI 2016 flight.
Of particular relevance to this work are the activation lines in the 1750--1850\,keV energy regime, given the desire to model background photopeaks near the signature \nuc{Al}{26} emission at 1809\,keV.
The simulations are conducted with a 12-detector mass model in order to account for all material in the COSI instrument.

\begin{figure}
\centering
\includegraphics[width=0.7\linewidth]{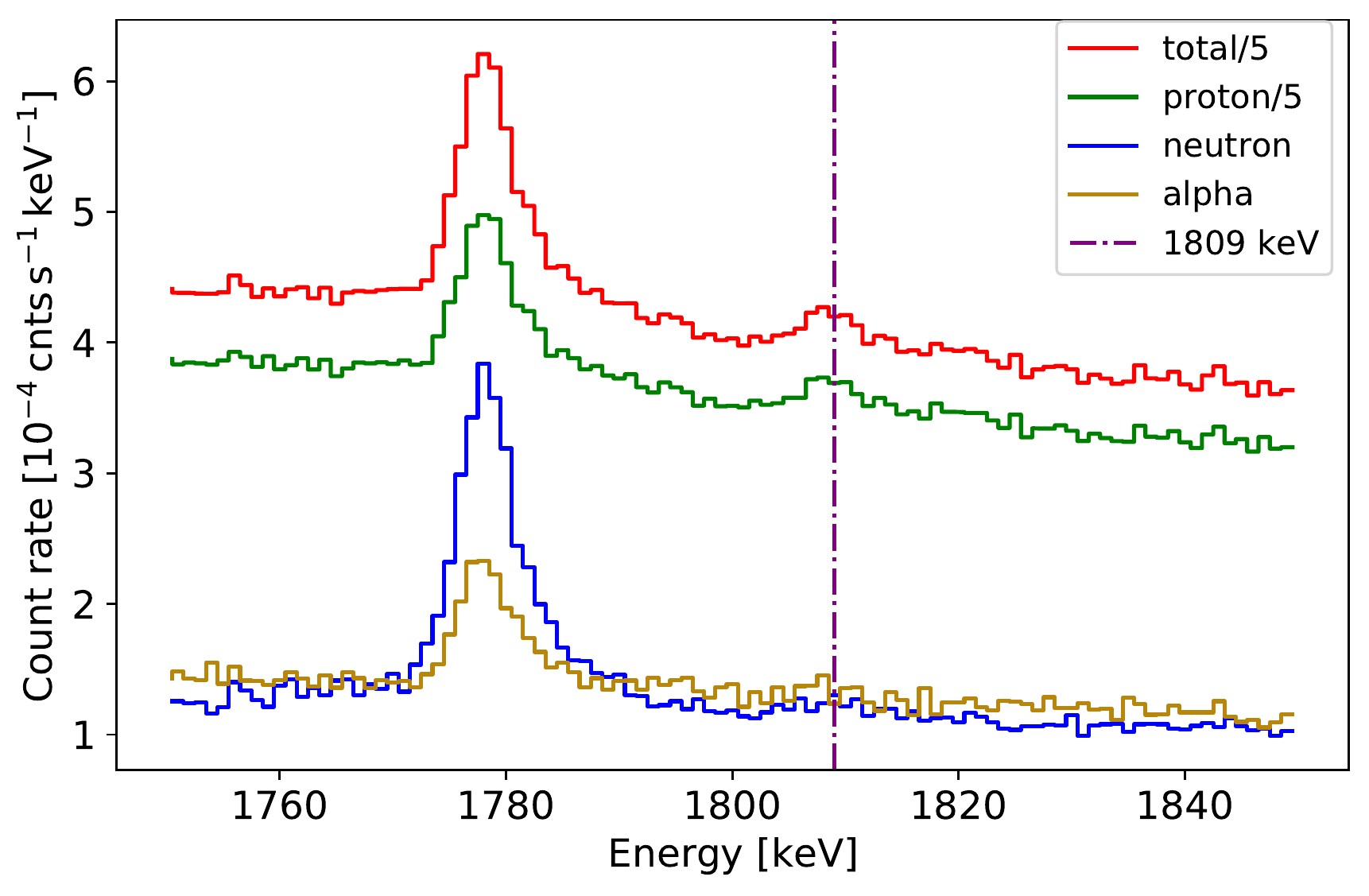}
\caption{Spectra of delayed emission from instrumental activation due to protons, neutrons, and $\alpha$-particles. The summed contribution of all components is shown in red. All Compton events between 1750--1850\,keV with Compton scattering angle between 0--90$^{\circ}$ are included.}
\label{fig:activation_46days_12dets_components}
\end{figure}

Spectra of the delayed emission, step (3), from each of the dominant particles are shown in Figure\,\ref{fig:activation_46days_12dets_components}.
Only limited event selections are applied to the data:
we show Compton events from all times between 1750--1850\,keV with Compton scattering angles from 0--90$^{\circ}$, no minimum distance between subsequent interactions, no Earth Horizon cut, and no pointing cut on the sky.
Additional cuts are used in the analysis to further restrict the events in this ``initial" data set to, for example, the signal and background regions (Sect.\,\ref{sec:data_selection}). 

Figure\,\ref{fig:activation_46days_12dets_components} shows that the protons constitute the large majority of activation background in the COSI 2016 flight.
The general shape of the activation spectra largely follows that seen in the spectrum of the background region flight data with minimal event selections (Figure\,\ref{fig:bg_model}).
The peaks at $\sim$1779\,keV and $\sim$1809\,keV are easily identifiable and their likely origins are documented in the literature as captures on $\mathrm{^{27}Al}$ (see Sect.\,\ref{sec:bg_model}).
The total count rates of both peaks, summed over particle type, are $\sim 3.0 \times 10^{-3}$ cnts keV$^{-1}$ s$^{-1}$ and $\sim 2.1 \times 10^{-3}$ cnts keV$^{-1}$ s$^{-1}$, comparable to those seen in Figure\,\ref{fig:bg_model} within an order of magnitude.

Notably absent from the activation spectra is the peak near 1764\,keV seen in Figure\,\ref{fig:bg_model} from the real flight background.
The literature widely attributes this line to the decay of $\mathrm{^{238}U}$ in instrument materials, and because this is a natural decay rather than a signature of de-excitation after activation of instrument materials, its absence from the instrumental activation simulation might be expected.
However, the true origin of this line in the real flight background remains uncertain.
Hence we employ an empirical description of the flight background which accounts for this line regardless of origin.

\subsection{Atmospheric Simulations}
\label{sec:appx_ling_sims}
Atmospheric $\gamma$-rays pose an enormous problem for balloon-borne instruments.
Susceptible to the glow of $\gamma$-rays from the Earth's atmosphere below the floating instrument, balloon-borne experiments must develop robust methods of rejecting atmospheric background.
Many instruments, including COSI, adopt anti-coincidence shielding to reject events emanating from below the gondola that are coincident with events in the germanium detectors.
COSI also uses an ``Earth Horizon Cut" that rejects events incident greater than 90$^{\circ}$ from the instrument's zenith, which is always pointed upward.
However, these methods do not guarantee complete background rejection (e.g. small physical gaps between anti-coincidence shielding) and modeling of the atmospheric background is necessary to understand the contamination of flight data by atmospheric background.

The atmospheric $\gamma$-ray background model by \citet{ling_model} presents a description of the 0.3--10\,MeV energy range at geomagnetic latitude $\lambda = 40^{\circ}$.
It derives an isotropic, semi-empirical source function which models the production of $\gamma$-ray continuum and lines per unit air mass.
The continuum is produced largely by bremsstrahlung of primary and secondary cosmic-ray electrons, neutral pion decays, and the scattering of incident photons to lower energies.
The dominant discrete contribution is a strong 511\,keV electron-positron annihilation line.
Other line components resulting from particle captures and subsequent decays, for example, are also possible.
The intensity of photons with incident energy $E'$ and incident angle $\theta$ (measured from zenith) seen by a detector at atmospheric depth $h$ [g\,cm$^{-2}$], as measured from the top of the atmosphere, is given by 

\begin{multline}
    \frac{dF(E',h)}{d\Omega} = \Big(\int_r S(E',x)\rho(x) \mathrm{exp}\Big[-\int_0^r\mu(E')\rho(r)dr\Big]\frac{dr}{4\pi} \\
    + \frac{dF_c(E')}{d\Omega}\mathrm{exp}\Big[-\int_0^{\infty}\mu(E')\rho(r)dr\Big]\Big)\,\mrm{ph\,cm^{-2}\,s^{-1}\,sr^{-1}\,MeV^{-1},}
    \label{eq:ling_model}
\end{multline}

\noindent where $\rho(x)$ is the air density for depth $x$ and $\mu(E')$ is the mass absorption coefficient.
While \citet{ling_model} provides expressions for the source functions $S(E',x)$ for both the continuum and line contributions, in this work we adopt a description of air density and mass absorption coefficient $\mu(E')$ given by \citet{picone2002nrlmsise}.
We choose one day of the 2016 flight to represent the atmospheric conditions over the entire flight because background model simulations are computationally intensive.
Given that the focus of this analysis is \nuc{Al}{26} from the Inner Galaxy, a day with maximum exposure of the Galactic Plane, corresponding to negative Earth latitudes, is chosen.
The following flight conditions corresponding to 2016 May 22 00:00:00 UTC are fed to the NRLMSISE-00 atmospheric model \citep{nrlmsise-00}:
flight altitude = 33.6\,km, latitude = $-$56.2$^{\circ}$, longitude = 161$^{\circ}$.
The model returns the densities of atmospheric atomic and molecular oxygen and nitrogen, as well as helium, argon, and hydrogen in units of cm$^{-3}$, the total mass density in g\,cm$^{-3}$, and the atmospheric temperature in Kelvin for heights of 0--100\,km.

The background model simulation runs in MEGAlib, using the above atmospheric quantities and an orientation file as inputs.
The balloon orientations are required so that COSI is pointed to the correct Galactic coordinates which mimic the entire 2016 flight.
Five quantities define the orientation:
time, the longitude and latitude of COSI's \textit{x}-axis, and the longitude and latitude of COSI's \textit{z}-axis.
Here, the \textit{z}-axis defines the instrument's optical axis (zenith = 0), and the \textit{x}-axis defines its azimuthal rotation.

Given the orientations for the complete COSI 2016 flight, with all pointings weighted by exposure time, and the NRLMSISE-00 atmospheric conditions from 2016 May 22, we run the simulation and process it using 10- and 9-detector mass models.
Concatenating the 10- and 9-detector Ling model simulation thus yielded a representation of atmospheric background over the COSI 2016 flight. 
%

%% For this sample we use BibTeX plus aasjournals.bst to generate the
%% the bibliography. The sample631.bib file was populated from ADS. To
%% get the citations to show in the compiled file do the following:
%%
%% pdflatex sample631.tex
%% bibtext sample631
%% pdflatex sample631.tex
%% pdflatex sample631.tex

%\bibliography{references}{}
\bibliographystyle{aasjournal}

% below copied and pasted from .bbl file in cache

%% This command is needed to show the entire author+affiliation list when
%% the collaboration and author truncation commands are used.  It has to
%% go at the end of the manuscript.
%\allauthors

%% Include this line if you are using the \added, \replaced, \deleted
%% commands to see a summary list of all changes at the end of the article.
%\listofchanges

\end{document}